\documentstyle[12pt]{article}

\input{psfig}

\font\mybbs=msbm10 at 9pt
\def\bbs#1{\hbox{\mybbs#1}}
\font\mybb=msbm10 at 12pt
\def\bb#1{\hbox{\mybb#1}}
\def\e{{\rm e}}

\newcommand{\newsection}[1]
{\vspace{5mm}
\pagebreak[3]
\addtocounter{section}{1}
\setcounter{equation}{0}
\setcounter{subsection}{0}
\setcounter{footnote}{0}
\begin{flushleft}
{\large\bf \thesection. #1}
\end{flushleft}
\nopagebreak
\medskip
\nopagebreak}

\newcommand{\newsubsection}[1]{
 \vspace{5mm}
\pagebreak[3]
\addtocounter{subsection}{1}
\noindent{ \bf \thesubsection. #1}
\nopagebreak
\vspace{2mm}
\nopagebreak}

\newlength{\extraspace}
\newlength{\extraspaces}
\begin{document}

\addtolength{\baselineskip}{.8mm}

\thispagestyle{empty}

\begin{flushright}
{\sc OUTP}-97-57P\\
  hep-th/9710179\\
\hfill{  }\\
Revised Version\\
\hfill{  }\\
March 1998
\end{flushright}
\vspace{.3cm}

\begin{center}
{\Large\bf Mirror Maps in Chern-Simons Gauge Theory}\\[15mm]

{\sc Leith Cooper}\footnote{Email: {\tt l.cooper2@physics.oxford.ac.uk}}, {\sc
Ian I.\ Kogan}\footnote{Email: {\tt i.kogan1@physics.oxford.ac.uk}} {\sc and
Richard J.\ Szabo}\footnote{Email: {\tt r.szabo1@physics.oxford.ac.uk}} \\[2mm]
{\it Department of Physics -- Theoretical Physics\\ University of Oxford\\ 1
Keble Road, Oxford OX1 3NP, U.K.} \\[15mm]

\vskip 1.0 in

{\sc Abstract}

\begin{center}
\begin{minipage}{14cm}

We describe mirror symmetry in $N=2$ superconformal field theories in terms of
a dynamical topology changing process of the principal fiber bundle associated
with a topological membrane. We show that the topological symmetries of
Calabi-Yau sigma-models can be obtained from discrete geometric transformations
of compact Chern-Simons gauge theory coupled to charged matter fields. We
demonstrate that the appearence of magnetic monopole-instantons, which
interpolate between topologically inequivalent vacua of the gauge theory,
implies that the discrete symmetry group of the worldsheet theory is realized
kinematically in three dimensions as the magnetic flux symmetry group. From
this we construct the mirror map and show that it corresponds to the
interchange of topologically non-trivial matter field and gauge degrees of
freedom. We also apply the mirror transformation to the mean field theory of
the quantum Hall effect. We show that it maps the Jain hierarchy into a new
hierarchy of states in which the lowest composite fermions have the same
filling fractions.

\end{minipage}
\end{center}

\end{center}

\noindent

\vfill
\newpage
\pagestyle{plain}
\setcounter{page}{1}
\stepcounter{subsection}
\renewcommand{\footnotesize}{\small}

\newsection{Introduction}

One of the most interesting consequences of $N=2$ worldsheet supersymmetry in
string theory is the occurence of physically smooth spacetime topology changing
processes. They result from a string duality called `mirror symmetry'
\cite{mirror}--\cite{greene} in which two topologically distinct Calabi-Yau
compactifications give rise to identical physical models. Heuristically, as
the size of a given compactification is made small, huge curvature fluctuations
modify the fabric of spacetime leading to a change in topology. The mirror
transformation relating these two distinct geometrical formulations of the same
physical situation connects strong to weak sigma-model coupling constants and,
via a judicious choice of geometrical model, seemingly difficult physical
questions can be analysed with perturbative ease. Mirror symmetry also leads to
the notion of `quantum geometry', which is the appropriate modification of
standard, classical geometry to make it suitable for describing the spacetime
physics implied by string theory, and it greatly simplifies the problem of
computing the moduli space of $N=2$ superconformal field theories which form
the set of string vacua.

Several attempts have been made to provide a rigorous mathematical framework,
for example using algebraic geometry and toric geometry, for the duality in
Calabi-Yau moduli space implied by the existence of mirror manifolds.
Conversely, mirror symmetry has been applied as a tremendously powerful
calculational tool in these same branches of mathematics. In this paper we
shall describe the mirror map in terms of a dynamical topology changing process
acting on a principal fiber bundle over a 3-manifold $\cal M$. The main idea is
to exploit the intimate relationship between two-dimensional conformal field
theory and the three-dimensional topological field theory defined by the
Chern-Simons action
\begin{equation}
kS_{\rm CS}^{[G]}[A]=\frac{k}{4\pi}\int_{\cal
M}\mbox{Tr}\left(A\wedge dA+\mbox{$\frac{2}{3}$}A\wedge A\wedge
A\right)
\label{csaction}\end{equation}
where $A$ is a connection on a principal fiber bundle over $\cal M$ with
structure group $G$. The quantum field theory defined by (\ref{csaction}) is
equivalent to the Wess-Zumino-Novikov-Witten ({\sc wznw}) model at level
$k\in{\bb Z}$ defined on a compact Riemann surface $\Sigma$. This can be seen
at the level of the physical state space of (\ref{csaction}) on the product
manifold ${\cal M}=\Sigma\times{\bb R}$, which is naturally isomorphic to the
finite-dimensional space of conformal blocks of the {\sc wznw} model
\cite{witten}, or in a path integral approach whereby the theory
(\ref{csaction}) defined on a 3-manifold $\cal M$ with boundary $\partial{\cal
M}=\Sigma$ induces the chiral gauged {\sc wznw} model on $\partial\cal M$
\cite{mooreET}.

The key feature of this correspondence is that various algebraic properties of
two-dimensional conformal field theories on $\partial\cal M$ can be understood
geometrically and dynamically in the three-dimensional approach. For instance,
the $n$-point correlation functions of the conformal field theory can be
decomposed
\begin{equation}
\left\langle\prod_{i=1}^nV^{(R_i)}(z_i,\bar
z_i)\right\rangle=\left\langle\prod_{i=1}^nV_{\rm
L}^{(R_i)}(z_i)\right\rangle\left\langle\prod_{i=1}^nV_{\rm R}^{(R_i)}(\bar
z_i)\right\rangle
\label{CFTcorr}\end{equation}
in terms of products of left and right conformal blocks, where $V_{\rm
L}^{(R_i)}(z_i)$ and $V_{\rm R}^{(R_i)}(\bar z_i)$ are the holomorphic and
anti-holomorphic chiral vertex operators corresponding to the left-right
symmetric vertex operator $V^{(R_i)}(z_i,\bar z_i)$ in a representation $R_i$
of the group $G$. In the corresponding three-dimensional description we
consider the 3-manifold ${\cal M}=\Sigma\times[0,1]$ whose two boundaries
$\Sigma_{\rm L}$ and $\Sigma_{\rm R}$ are connected by a finite time interval.
A
Chern-Simons gauge theory in $\cal M$ induces both left- and right-moving
sectors of the two-dimensional conformal field theory, and an insertion of a
vertex operator on the worldsheet $\Sigma$ is equivalent to insertions of the
chiral vertex operators $V_{\rm L}^{(R)}(z)$ and $V_{\rm R}^{(R)}(\bar z)$
on the left- and right-moving worldsheets $\Sigma_{\rm L}$ and $\Sigma_{\rm
R}$, respectively. The insertions corresponding to the correlation functions
(\ref{CFTcorr}) are induced by path-ordered products of the open Wilson line
operators \cite{witten}--\cite{szaboET}
\begin{equation}
W_{{\cal C}_{z_1,\bar z_1};\dots;{\cal C}_{z_n,\bar
z_n}}^{(R_1,\dots,R_n)}[A^{(1)},\dots,A^{(n)}]=\prod_{i=1}^n~\mbox{Tr}~P\exp
\left(i\int_{{\cal C}_{z_i,\bar z_i}}A^{(i)a}R_i^a\right)
\label{wilsonvert}\end{equation}
along the oriented paths ${\cal C}_{z_i,\bar z_i}\subset{\cal M}$ with
endpoints $z_i\in\Sigma_{\rm L}$ and $\bar z_i\in\Sigma_{\rm R}$.
Correlators of insertions of the Wilson lines (\ref{wilsonvert}) in $\cal M$
induce phase factors from adiabatical rotation of charged particles coupled to
the Chern-Simons gauge fields $A^{(i)}$ in the representations $R_i$. The
quantum particles propagate along ${\cal C}_{z_i,\bar z_i}$ from left- to
right-moving worldsheets, so that the corresponding linkings of the Wilson
lines
from the adiabatical rotations in $\cal M$ are equivalent to braidings of the
associated vertex operators on $\Sigma$.

The induced phases arising from the braiding operations are given by the
Knizhnik-Zamolodchikov formula \cite{KZ}
\begin{equation}
\Delta_R(G_k)=\frac{T_R(G)}{k+C_2(G)}
\label{KZspin}\end{equation}
for the anomalous scaling dimensions of primary operators in the corresponding
current algebra $G_k$ based on $G$ at level $k$, where $T_R(G)$ is the
quadratic Casimir of the representation $R$ and $C_2(G)$ is the quadratic
Casimir of the adjoint representation of $G$. In the three-dimensional approach
the weights (\ref{KZspin}) can be obtained as the Aharonov-Bohm phases that
arise from adiabatical rotation of charged particles, interacting with a
Chern-Simons gauge field, about one another \cite{anyon}, or equivalently from
the corresponding Aharonov-Bohm scattering amplitudes between dynamical charged
particles \cite{szaboET}. The key property behind this equivalence is that a
particle of charge $q$ minimally coupled to a Chern-Simons gauge theory is also
a source of magnetic flux $\Phi_q$. For an abelian gauge field $A$, this
follows from the Gauss law
\begin{equation}
\frac k{4\pi}B\equiv\frac k{4\pi}\,\epsilon^{0ij}\partial_iA_j=J^0
\label{gausslaw}\end{equation}
obtained by varying the action $kS_{\rm CS}^{[U(1)]}[A]$ minimally coupled to
the particle current $J^\mu$, with respect to the temporal component $A_0$ of
$A$. From (\ref{gausslaw}) we see that the charge $q$ also carries magnetic
flux
\begin{equation}
\Phi_q\equiv\frac1{2\pi}\int_Dd^2x~B(x)=\frac{2q}k
\label{flux}\end{equation}
where $D\subset\cal M$ is a disc in a neighbourhood of the charged matter.
The conformal weights (\ref{KZspin}) are therefore induced spins that are the
Aharonov-Bohm phases from adiabatical rotation of particles of charge
$q=\sqrt{T_R(G)}$, i.e.
\begin{equation}
\Psi(\e^{2\pi i}(x_1-x_2))=\e^{4\pi i\Delta_R(G_k)}~\Psi(x_1-x_2)=\e^{2\pi
iq\Phi_q}~\Psi(x_1-x_2)
\label{2partwave}\end{equation}
where $\Psi$ is the 2-particle wavefunction.

In \cite{us} these properties were exploited to obtain a dynamical description
of models representing a particular region of the moduli space of $N=2$
superconformal field theories. The $N=2$ superconformal algebra is generated by
the usual Virasoro stress-energy tensor $(T(z),\bar T(\bar z))$, an extra
$U(1)$ current $(J(z),\bar J(\bar z))$ of conformal dimension 1 (which is not
present in the $N=0$ and $N=1$ algebras), and two fermionic supercurrents
$(G^\pm(z),\bar G^\pm(\bar z))$ with $U(1)$ charges $\pm1$. Actually, there is
a one-parameter family of supercurrents labelled by their various boundary
conditions, which continuously interpolates among a family of isomorphic $N=2$
superconformal algebras. This algebraic property of the conformal field theory
was described dynamically in \cite{us} in terms of the propagation of charged
particles coupled to the relevant Chern-Simons gauge theory. The supercurrents
also define non-singular and closed finite-dimensional chiral rings in the
operator product algebra of primary fields \cite{mirror}. A primary field lies
in the chiral-chiral ring if it has non-singular operator product with
$(G^+(z),\bar G^+(\bar z))$, and in the chiral-antichiral ring if its operator
product with $(G^+(z),\bar G^-(\bar z))$ is regular.

\begin{figure}[htb]
\centerline{\psfig{figure=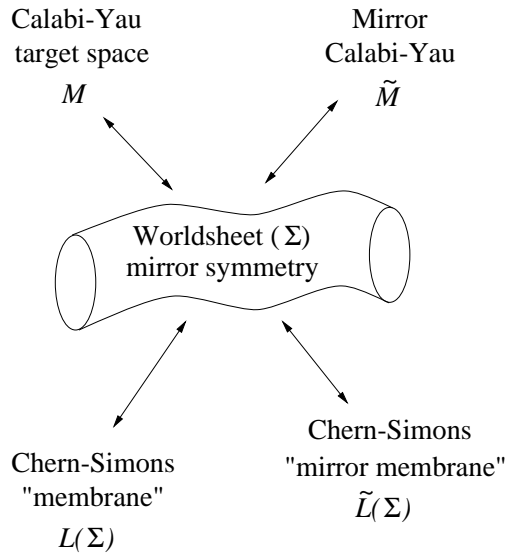,height=0.45\textwidth}}
\small
\caption{\baselineskip=12pt In string theory, worldsheet mirror symmetry
provides a map $M\leftrightarrow\widetilde{M}$ between two Calabi-Yau target
spaces with different topologies. In the topological membrane approach,
worldsheet mirror symmetry provides a map
$L(\Sigma)\leftrightarrow\widetilde{L}(\Sigma)$ between two Chern-Simons gauge
theories with different topological line bundles.}
\label{example}\end{figure}

In the following we will describe, within a geometric setting, some dynamical
properties of the chiral rings of $N=2$ superconformal field theories on
generic compact Riemann surfaces. We will show how the topological symmetries
of Calabi-Yau manifolds can be realized by geometrical transformations on the
3-manifold $\cal M$ with respect to the coupling of charged matter to a
Chern-Simons gauge theory. Such a matter-coupling represents a deformation of
the corresponding conformal field theory. The ``mild" symmetries, such as Hodge
duality and K\"ahler symmetry, arise from orientation-reversing isometries of
$\cal M$, such as parity and time-reversal. We will show that mirror symmetry
arises when one makes the topology of the gauge theory non-trivial. Namely,
when the gauge group is compact, magnetic monopole-instanton induced
transitions alter the topology of the given principal fiber bundle by changing
its first Chern class. In this case the discrete symmetry group of the
superconformal model in the given region of moduli space is realized
kinematically as the magnetic flux symmetry group of the gauge theory. The
Hilbert space of the topologically non-trivial Chern-Simons theory provides a
representation of the chiral rings of the corresponding conformal field
theories. Mirror symmetry is then a mapping between two non-perturbative
dynamical processes in which topologically non-trivial matter field
configurations are exchanged with topologically non-trivial configurations of
the gauge fields that interpolate between inequivalent vacua of the gauge
theory. We also show how these maps are related to `topological duality'
transformations that are reminescent of $T$-duality transformations in string
theory. These results show that, within the topological membrane approach
\cite{tm}, in which the string theory is studied by filling in the worldsheet
and viewing it as the boundary of a 3-manifold $\cal M$, mirror symmetry is
induced by a non-perturbative dynamical process that relates two topologically
distinct membranes to one another, both of which induce the {\it same} string
theory on $\partial{\cal M}$. This process is illustrated schematically in fig.
\ref{example}. In certain instances this maps a topologically
non-trivial fiber bundle onto a trivial one, so that, from a mathematical
perspective, the symmetry of the worldsheet theory implies that difficult
questions concerning a principal fiber bundle can be enormously simplified by
mapping it onto its mirror.

\bigskip

\bigskip

As a more concrete realization of these formalisms, we consider the most
successful physical application of Chern-Simons gauge theory in condensed
matter physics, namely the quantum Hall effect \cite{qhecs}. A quantum Hall
system is a two-dimensional gas of electrons interacting with a uniform
neutralizing background potential and with a strong perpendicular magnetic
field. In certain instances, the transverse conductivity exhibits plateaus at
certain values of the (fractional) number of filled Landau levels. An effective
(2 + 1)-dimensional description of these phenomena is provided by coupling an
additional (fictitious) gauge field to the electric charge current and adding
to the Lagrangian a corresponding Chern-Simons term. The mirror map applied to
these systems is especially transparent and we present transformation laws
which map a given quantum Hall system onto its `mirror'. The Hall conductivity
is changes by a simple rescaling, and, in certain cases, it is invariant under
this transformation indicating a true `mirror symmetry' of quantum Hall
systems. We discuss some noteworthy physical aspects of such mappings, such as
the role of the topologically non-trivial field configurations in these
instances, which could have bearings on experimental situations. In this way we
show that not only is the phenomenon of mirror symmetry dynamical in origin,
but it also exhibits universal applications in other branches of physics and
mathematics.

The outline of this paper is as follows. In section 2 we briefly review the
basic ideas of mirror symmetry and the construction of mirror manifolds via
algebraic isomorphisms of $N=2$ superconformal minimal models. In section 3 we
describe the relevant superconformal field theories in terms of matter-coupled
three-dimensional gauge theory and give the discrete geometric transformations
of $\cal M$ which represent the symmetries of the Calabi-Yau spaces. In section
4 we describe the properties of compact Chern-Simons theory, emphasizing the
appearence of monopole-instanton field configurations as objects inducing
topology change on the principal fiber bundle of the gauge theory. Then in
section 5 we present the mirror map in full detail in terms of charge
deformations represented by Wilson line operators, and also by a detailed
construction of the Hilbert space of the matter-coupled topological field
theory. In section 6 we apply these constructions to the quantum Hall effect,
and finally section 7 contains some concluding remarks.

\newsection{Mirror maps in $N=2$ superconformal field theories}

In this section we shall briefly review some of the main ideas behind mirror
symmetry in $N=2$ superconformal field theories that will be used throughout
this paper. A more detailed review can be found in \cite{greene}.

\newsubsection{Marginal operators and mirror symmetry}

A deformation of a conformal field theory with action $S_{\rm CFT}$ by some
operators ${\cal O}_i$ (in the original theory) of conformal dimensions
$(\Delta_i,\bar\Delta_i)$ is described by an action of the form
\begin{equation}
S[\mu]=S_{\rm CFT}+\sum_i\lambda_i[\mu]\int_{\Sigma}d^2z~{\cal O}_i(z,\bar z)
\label{cftdeform}\end{equation}
where $\lambda_i[\mu]$ are coupling constants which depend on a scale $\mu$.
The corresponding renormalization group flows allow one to smoothly interpolate
between various two-dimensional renormalizable quantum field theories and the
model with action $S_{\rm CFT}$ considered as infrared or ultraviolet fixed
points of the flows \cite{cardy}. Of central importance to understanding the
structure of the moduli space of $N=2$ superconformal field theories are
deformations by marginal operators, i.e. those with scaling dimensions
$\Delta+\bar\Delta=2$. These operators deform a given conformal field theory to
a ``nearby" one of the same central charge $c$ and thereby generate a family of
isomorphic conformal field theories which are all continuously related to each
other. The subset of marginal operators with conformal weights
$(\Delta,\bar\Delta)=(1,1)$ that continue to have dimensions $(1,1)$ after
perturbation by any other operator in the collection are said to be truly
marginal. Such fields naturally span the tangent space to any point in the
moduli space and can be used to move around the moduli space without spoiling
conformal invariance.

An abstract $N=2$ superconformal field theory with $c=3d$ has two types of
truly marginal operators, ${\cal O}_{(1,-1)}$ and ${\cal O}_{(1,1)}$, which are
constructed out of primary fields in the chiral-antichiral ring with $U(1)$
charges $({\cal Q},\bar{\cal Q})=(1,-1)$ and in the chiral-chiral ring with
$U(1)$ charges $(1,1)$, respectively (see for example \cite{aspinwall,greene}).
They therefore differ only by the sign of a $U(1)$ charge in the
anti-holomorphic sector. These operators can be given a geometrical
interpretation in terms of a non-linear sigma-model on a Calabi-Yau target
space $M$ of complex dimension $d=\frac c3$. The operator ${\cal O}_{(1,-1)}$
corresponds to deformations of the K\"ahler structure of $M$ which, to lowest
order, can be represented in terms of a harmonic $(1,1)$-form on $M$. The
operator ${\cal O}_{(1,1)}$, on the other hand, corresponds to deformations of
the complex structure of $M$  which can be represented in terms of a harmonic
$(1,d-1)$-form on $M$. Both deformations preserve the Calabi-Yau conditions on
$M$ and so the $N=2$ superconformal invariance of the original non-linear sigma
model is maintained. The chiral-chiral ring is thus a deformation of the DeRham
cohomology ring of the Calabi-Yau manifold $M$. For the chiral-chiral primary
subspaces with $U(1)_{\rm L}\times U(1)_{\rm R}$ charges \cite{mirror}
\begin{equation}
\left({\cal Q}^{(i)}~,~\bar{\cal Q}^{(j)}\right)=\left(\mbox{$i-\frac
c6~,~\frac c6-j$}\right)
\label{ijcharges}\end{equation}
the dimensions of these subspaces are equal to the Hodge numbers
$h^{i,j}(M)=\dim_{\bbs C}H^{i,j}(M)$, the dimensions of the spaces of harmonic
$(i,j)$-forms on $M$. In fact, for the sigma-model with target space $M$, the
chiral-chiral ring is the cohomology ring of $M$ deformed by instanton effects
(i.e. complex curves in $M$) \cite{witteninst}.

In terms of the abstract conformal field theory, the distinction between the
two types of truly marginal operators ${\cal O}_{(1,-1)}$ and ${\cal
O}_{(1,1)}$ is rather trivial, being just the sign of a conventional $U(1)$
charge. The $N=2$ superconformal algebra is invariant under reflection of the
corresponding current $\bar J(\bar z)$. On the other hand, their geometrical
counterparts (the Dolbeault cohomology groups $H^{1,1}(M)$ and $H^{1,d-1}(M)$,
respectively) differ far more significantly. This led to the conjecture
\cite{mirror} that to each Calabi-Yau manifold $M$ there corresponds a {\em
mirror} Calabi-Yau manifold $\widetilde{M}$ corresponding to the same conformal
field theory but with ${\cal O}_{(1,-1)}\in H^{1,d-1}(\widetilde{M})$ and
${\cal O}_{(1,1)}\in H^{1,1}(\widetilde{M})$. Each type of marginal operator
would then correspond to either a K\"ahler or a complex structure deformation
of the mirror pair $(M,\widetilde{M})$. Geometrically, this means that the
Hodge numbers of $M$ are related to those of $\widetilde{M}$ by
\begin{equation}
h^{1,1}(M)=h^{1,d-1}(\widetilde{M})~~~~~~,~~~~~~h^{1,d-1}(M)=
h^{1,1}(\widetilde{M})
\end{equation}
This correspondence generalizes to the other Hodge numbers as well
\cite{mirror}. Now the chiral-antichiral ring is a deformed cohomology ring of
the mirror Calabi-Yau manifold $\widetilde{M}$ with the Hodge numbers
\begin{equation}
h^{i,j}(\widetilde{M})=h^{i,d-j}(M)
\label{hijmirror}\end{equation}
and so the Hodge diamond for $\widetilde{M}$ is mirror reflection through a
diagonal axis of the Hodge diamond for $M$. Furthermore, the chiral-chiral ring
coincides with the undeformed Dolbeault cohomology ring with values in the
exterior algebra of the holomorphic tangent bundle of $\widetilde{M}$. In this
way, one can extract non-trivial information about instanton numbers for a
Calabi-Yau space $M$ from a simple calculation of the Dolbeault cohomology on
the mirror image $\widetilde{M}$ of $M$ (see \cite{aspinwall} for some
examples).

\newsubsection{Constructing the mirror manifold}

The key to constructing the mirror manifold, which was first carried out in
\cite{mirrors}, of a conformal field theory $\bf C$
associated with a Calabi-Yau manifold $M$ (or, more generally, the equivalence
class of conformal field theories which are isomorphic to the non-linear
sigma-model with target space $M$) is to find an operation which flips the sign
of the right-moving $U(1)_{\rm R}$ charge of each marginal operator in $\bf C$
(compare (\ref{ijcharges}) and (\ref{hijmirror})). This operation maps $\bf C$
to an isomorphic conformal field theory $\widetilde{\bf C}$ (with corresponding
Calabi-Yau space $\widetilde{M}$) such that ${\cal O}_{(1,-1)}\in
H^{1,1}(M)\rightarrow\widetilde{\cal O}_{(1,1)}\in H^{1,d-1}(\widetilde{M})$
and ${\cal O}_{(1,1)}\in H^{1,d-1}(M)\rightarrow\widetilde{\cal O}_{(1,-1)}\in
H^{1,1}(\widetilde{M})$. Since $\bf C\cong\widetilde{\bf C}$, the marginal
operator ${\cal O}_{(1,-1)}$ (and similarly ${\cal O}_{(1,1)}$) can be thought
of as either a K\"ahler (complex structure) deformation on $M$ or as a complex
structure (K\"ahler) deformation on $\widetilde M$. This observation typically
simplifies the computation of the moduli space of complex structures of a
Calabi-Yau manifold $M$ (and hence of the conformal field theory $\bf C$)
\cite{aspinwall}.

The $N=2$ superconformal {\it minimal} models do admit such an operation,
orbifolding, which yields an isomorphic conformal field theory related to the
original one by a change in the sign of all $U(1)_{\rm R}$ charges. Then, by
transporting the orbifold operation on these minimal models to the region of
moduli space corresponding to a Calabi-Yau sigma-model, this gives an explicit
method for constructing the mirror manifold \cite{witteninst,minmirror}. So an
important ingredient in the construction of mirror manifolds is an
understanding of the $N=2$ minimal model conformal field theories.

As described in \cite{greene} (see also \cite{us}), the $k$-th $N=2$
superconformal minimal model is isomorphic to the coset
\begin{equation}
M_k\cong SU(2)_k\times SO(2)_2/U(1)_{k+2}
\label{minimal}\end{equation}
of ordinary $N=0$ {\sc wznw} models. Heuristically, it represents the operation
of removing a free boson at one radius by dividing out the $U(1)$ subgroup and
putting back a free boson at a different radius. The $SO(2)_2$ part of the
coset (\ref{minimal}) can be used to represent the extra $U(1)$ current
$(J(z),\bar J(\bar z))$ and, upon fermionization, also the fermion fields of
the $N=2$ superconformal algebra. The Virasoro central charge of
(\ref{minimal}) is
\begin{equation}
c_k=\frac{3k}{k+2}
\label{centralk}\end{equation}
The $N=2$ superconformal primary fields are labelled (in part) by their
conformal dimensions $(\Delta_{j,m},\bar\Delta_{\bar j,\bar m})$ and also by
the extra $U(1)$ charges $({\cal Q}_m,\bar{\cal Q}_{\bar m})$. In the
holomorphic Neveu-Schwarz sector of the theory they are given by
\begin{equation}
\Delta_{j,m}=\frac{j(j+1)}{k+2}-\frac{m^2}{k+2}~~~~~~,~~~~~~{\cal
Q}_m=-\frac{2m}{k+2}
\label{quants}\end{equation}
where $m=-j,-j+1,\dots,j-1,j$ are the magnetic quantum numbers of a spin-$j$
representation of $SU(2)$. Unitarity of this representation imposes an upper
bound on the isospin quantum numbers as
\begin{equation}
\mbox{$j=0,\frac12,1,\frac32,\dots,\frac k2$}
\label{jupper}\end{equation}

The $N=2$ minimal model is invariant under a \mbox{${\bb Z}_2\times{\bb
Z_{k+2}}\times \bar{{\bb Z}}_2\times\bar{{\bb Z}}_{k+2}$}
discrete symmetry group. The ${\bb Z}_{k+2}$ part of this symmetry group arises
from the constraint (\ref{jupper}) which implies that the charges ${\cal Q}_m$
are cyclic with period $k+2$, and it acts on the superconformal primary fields
by the conformal group element $\e^{2\pi iJ_0}$. It is essentially the discrete
symmetry group of level $k+2$ parafermions \cite{zamfat}. The ${\bb Z}_2$
symmetry is a charge conjugation symmetry represented by the fermion number $s$
which is defined modulo 4. The orbifold $M_k/{\bb Z}_{k+2}$ is a new conformal
field theory $\widetilde{M_k}$ which is isomorphic to $M_k$
\cite{zamfat,gepnerET1987a}. Moreover, the map from $\widetilde{M_k}$ to $M_k$
is simply $m\to-m$, which changes the sign of the $U(1)_{\rm R}$ charge
associated with each (anti-holomorphic) primary field. This is precisely the
map required to construct the corresponding Calabi-Yau mirror manifold. In what
follows we shall describe this mirror map from the point of view of
Chern-Simons gauge theory.

\newsection{Three-dimensional description of $N=2$ minimal models and
topological symmetries}

In \cite{us} it was shown that the $N=2$ minimal models (\ref{minimal}) can be
described completely using three independent Chern-Simons gauge fields $A$, $B$
and $C$ with the gauge field action
\begin{equation}
{\cal I}_k[A,B,C]=kS_{\rm CS}^{[SU(2)]}[A]+2S_{\rm CS}^{[SO(2)]}[B]
-(k+2)S_{\rm CS}^{[U(1)]}[C]
\label{abc}\end{equation}
The basic observables (\ref{quants}) of $M_k$ can be described in
three-dimesional terms by coupling the action (\ref{abc}) to charged matter. An
important feature of this, in the context of the last section, is that the
addition of charged matter to the bulk $\cal M$ will induce a deformation of
the two-dimensional conformal field theory on the boundary $\Sigma$
\cite{kogan2,kogan3}. In critical string theory the statistical sum of a
deformed conformal field theory with action (\ref{cftdeform}) gives a
generating function for correlators in an external field. In the case of a
single charge deformation it is given by
\begin{eqnarray}
{\cal Z}&=&\int D\phi~\exp\left(-S_{\rm CFT}[\phi]+\lambda\int_\Sigma
d^2z~V(z,\bar z)\right)\nonumber
\\&=&1+\sum_{n=1}^\infty\frac{\lambda^n}{n!}\int_\Sigma
d^2z_1~\cdots\int_\Sigma d^2z_n~\left\langle V(z_1,\bar z_1)\cdots V(z_n,\bar
z_n)\right\rangle
\label{statsum}\end{eqnarray}
where the averages denote $n$-point correlation functions in the unperturbed
conformal field theory. These correlators coincide with the expectation values
of Wilson line operators (\ref{wilsonvert}) in the corresponding Chern-Simons
gauge theory. A gas of open Wilson lines (\ref{wilsonvert}) therefore describes
charged matter in $\cal M$ corresponding to a deformation of the
two-dimensional conformal field theory.

The coupling of charged matter to the Chern-Simons gauge theory introduces
propagating degrees of freedom in the bulk and thus ruins the topological
property of the three-dimensional field theory. Furthermore, the deformation
parameters $\lambda_i[\mu]$ of the induced conformal field theory are functions
of the parameters of the matter fields. Thus a (truly marginal) deformation can
be described dynamically by varying the parameters of the charged matter in
three dimensions (for example the chemical potentials). These ideas were
explored in detail for the $N=0$ and $N=1$ minimal models in \cite{kogan1}.

The coupling of matter to the action (\ref{abc}) is therefore an essential
ingredient for the construction of the mirror map in three-dimensional terms.
As described in \cite{us}, the first step in obtaining a three-dimensional
description of the minimal models (\ref{minimal}) is to minimally couple the
$SU(2)$ gauge field $A$ in (\ref{abc}) to a matter current
$J_a^{(j)\mu}R^{(j)a}$ in a spin-$j$ representation $R^{(j)}$ of $SU(2)$. Then
the induced spin of this charged matter is $j(j+1)/(k+2)$ (see (\ref{KZspin})).
We also minimally couple the $U(1)$ Chern-Simons gauge field $C$ at level
$-(k+2)$ to a current $J^{(m)\mu}$ carrying an abelian charge $q=m$
corresponding to the magnetic quantum numbers of this same spin-$j$
representation. Then the total induced spin $\Delta_{j,m}$ of the
matter-coupled action
\begin{equation}
{\cal I}_k^{(j,m)}[A,B,C]={\cal I}_k[A,B,C]+\int_{\cal
M}\left(2j(j+1)A_\mu^aJ_a^{(j)\mu}+C_\mu J^{(m)\mu}\right)
\label{Imatter}\end{equation}
coincides precisely with the conformal dimensions in (\ref{quants}) for the
$N=2$ minimal model in the holomorphic Neveu-Schwarz sector. Furthermore, the
abelian magnetic flux (\ref{flux}) carried by the charge $q=m$ coupled to the
gauge field $C$ (at level $-(k+2)$) coincides precisely with the $U(1)$ charges
in (\ref{quants}), $\Phi_m={\cal Q}_m$.\footnote{See \cite{us} for a precise
description of the relationship between the abelian flux $\Phi_m$ and the total
$U(1)$ charge of the conformal group generator $J(z)$.}

The discrete symmetries of $M_k$ described in the previous section can be
explained dynamically in the three-dimensional picture. First of all, the
$U(1)_{\rm R}$ sector of the minimal model depends on the fermion number $s$
and orbifolding by ${\bb Z}_2$ sends $s\to-s$ which is related to the {\sc gso}
projection \cite{greene1996a}. This projection was described in detail in
\cite{us} in terms of the propagation of an extra charge coupled to the
$SO(2)_2$ gauge field $B$ in (\ref{abc}). This field has the effect of mapping
charged bosons into fermions and vice-versa through the Aharonov-Bohm effect
(see (\ref{2partwave})), and as such it represents the supercharges of the
$N=2$ supersymmetric theory. This therefore yields a dynamical picture of the
discrete ${\bb Z}_2$ symmetry of $M_k$. As for the parafermionic symmetry of
$M_k$, from (\ref{2partwave}) we likewise see that it is an {\it anyonic}
symmetry of the matter-coupled Chern-Simons theory, i.e. that the charged
particles acquire fractional exchange statistics parametrized by the phases
$\e^{2\pi im\Phi_m}\in{\bb Z}_{k+2}$. From a quantum mechanical perspective,
all physical quantities are invariant under transformation of the wavefunctions
by these statistical phases. In the next section we will see how this
statistical exchange symmetry arises from a kinematical property of the pure
gauge theory.

Moreover, we see that left- and right-moving $U(1)$ charges (which from
(\ref{ijcharges}) are associated with the Hodge numbers $h^{i,j}(M)$) not being
equal requires $\Phi_{\rm L}\neq\Phi_{\rm R}$, i.e. charge non-conservation as
the $U(1)$ particle propagates from left to right boundaries of ${\cal
M}=\Sigma\times[0,1]$. Exactly this type of process arises in {\it compact}
Chern-Simons gauge theory \cite{monoinst,kaiming} whereby a monopole-instanton
transition can change the charge of a particle along a Wilson trajectory. In
particular, the mirror map $m\to-m$ which changes the sign of the $U(1)_{\rm
R}$ charge in the $N=2$ minimal model (\ref{minimal}) corresponds in
three-dimensional terms to flipping the sign of the charge $q=m$ coupled to the
abelian Chern-Simons gauge field $C$. Thus the key to understanding the mirror
map $m\to-m$ from a three-dimensional perspective lies in the nature of the
gauge theory of the field $C$. This will be the topic of the next section.

\begin{figure}[htb]
\centerline{\psfig{figure=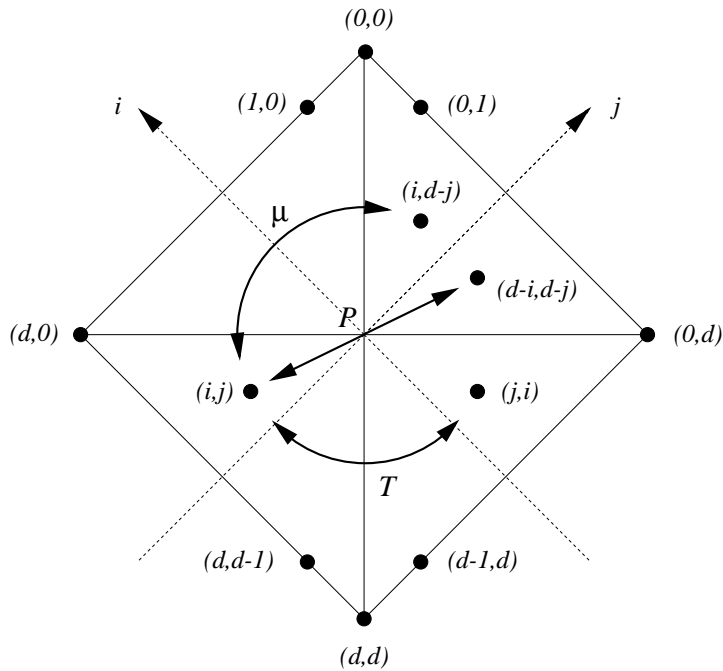,height=0.55\textwidth}}
\small
\caption{\baselineskip=12pt Symmetries of the Hodge diamond correspond to
discrete geometric transformations of the three-dimensional gauge theory. Each
entry in the diamond represents a Hodge number $h^{i,j}$. Hodge duality and the
K\"ahler condition can be viewed as a parity transformation $P$ and
time-reversal $T$, respectively, of the 3-manifold $\cal M$. Mirror symmetry
$\mu$ is a reflection about the main diagonal of the Hodge diamond and is
described in the three-dimensional picture by a non-perturbative dynamical
process.}
\label{diamond}\end{figure}

An interesting aspect of the three-dimensional description is that it provides
full geometric pictures for all of the topological symmetries of the associated
Calabi-Yau space $M$ represented by its Hodge diamond. For instance, Hodge
duality of the DeRham cohomology ring can be represented through the K\"ahler
condition as
\begin{equation}
H^n(M)\cong\bigoplus_{i+j=n}H^{i,j}(M)\cong H^{2d-n}(M)
\label{hodgeduality}\end{equation}
Comparing the Dolbeault cohomology groups on the left-hand side of the second
isomorphism in (\ref{hodgeduality}) with those of the right-hand side, and
using complex conjugation along with the K\"ahler condition, we find the
equalities
\begin{equation}
h^{d-i,d-j}(M)=h^{i,j}(M)~~~~~~,~~~~~~h^{i,j}(M)=h^{j,i}(M)
\label{dualhij}\end{equation}
among the various Hodge numbers of $M$. The first set of equalities in
(\ref{dualhij}) represents a parity transformation $P:{\cal M}\to{\cal M}^*$ of
the three-dimensional spacetime (which changes its orientation) under which the
magnetic fluxes transform as pseudo-scalars, i.e.
\begin{equation}
\left(\Phi_m~,~\bar\Phi_{\bar m}\right)~{\buildrel
P\over\longrightarrow}~\left(-\Phi_m~,~-\bar\Phi_{\bar m}\right)
\label{paritymap}\end{equation}
This property follows from (\ref{ijcharges}) which implies that
\begin{equation}
\left(-\Phi_m^{(i)}~,~-\bar\Phi_{\bar m}^{(j)}\right)=\left(\mbox{$\frac
d2$}-i~,~j-\mbox{$\frac d2$}\right)=\left((d-i)-\mbox{$\frac
d2$}~,~\mbox{$\frac d2$}-(d-j)\right)=\left(\Phi_m^{(d-i)}~,~\bar\Phi_{\bar
m}^{(d-j)}\right)
\label{parityexpl}\end{equation}
The Chern-Simons action is parity-odd and this transformation can be absorbed
in a reflection $(k+2)\to-(k+2)$ of the coefficient of the gauge field $C$ in
(\ref{abc}). The second set of equalities relate a charge non-conservation
process $q_i\to q_j$ as the charge propagates from left- to right-moving
worldsheets on ${\cal M}=\Sigma\times[0,1]$ to the same non-conservation on
propagation from right- to left-moving sectors of $\Sigma$, i.e.
\begin{equation}
\left(\Phi_m~,~\bar \Phi_{\bar m}\right)~{\buildrel
T\over\longrightarrow}~\left(\bar\Phi_{\bar m}~,~\Phi_m\right)
\label{timerevmap}\end{equation}
This is achieved by reversing the orientation of the worldlines of the
particles in the Wilson lines (\ref{wilsonvert}), i.e. a time-reversal
transformation $T:{\cal M}\to{\cal M}^*$, under which the Chern-Simons term is
again odd. Hodge duality and the K\"ahler structure of $M$ in the
three-dimensional picture is thus represented by discrete orientation-reversing
isometries of the 3-manifold $\cal M$. Thus, in addition to the mirror
reflection symmetries across the diagonals that will be described in the
following, the geometrical and dynamical properties of the three-dimensional
quantum field theory (\ref{Imatter}) also account in a simple way for the
horizontal and vertical symmetries (\ref{dualhij}) of the Hodge diamond of $M$
(see fig. \ref{diamond}).

\bigskip

\bigskip

\newsection{Topology of compact Chern-Simons gauge theory}

The mirror map exchanges the marginal deformation operators ${\cal O}_{(1,1)}$
and ${\cal O}_{(1,-1)}$. The latter deformation represents the propagation of a
charge $q$ coupled to the gauge field $C$ in (\ref{abc}) which flips its sign
as it propagates through the bulk $\cal M$ from $\Sigma_{\rm L}$ to
$\Sigma_{\rm R}$. As we will now explain, such a charge non-conservation
process is non-perturbative in character and can be accounted for by a
non-trivial topology for the Chern-Simons gauge theory. The exchange of the two
marginal deformations will then correspond to the interchange of topologically
trivial and non-trivial gauge propagations represented by the Wilson lines
(\ref{wilsonvert}).

\newsubsection{Topology change of the principal fiber bundle}

Consider the $U(1)_{k+2}$ Chern-Simons gauge field $C$. It is a connection of a
complex line bundle $L\to{\cal M}$ over the 3-manifold $\cal M$ with curvature
$F(C)=dC$. If we consider the product manifold ${\cal M}=\Sigma\times[0,1]$,
with $\Sigma$ a compact Riemann surface of genus $g$, then for each fixed time
$t\in[0,1]$ the field equations in the absence of sources are $B\equiv
F(C|_\Sigma)=0$ (see (\ref{gausslaw})). The topological quantum field theory
thus localizes onto the moduli space of flat gauge connections modulo gauge
transformations on the Riemann surface $\Sigma$. From this fact it is natural
to restrict the line bundle $L\to{\cal M}$ to a line bundle $L_\Sigma\to\Sigma$
with curvature the magnetic field $B$. It is classified topologically by its
first Chern characteristic class
\begin{equation}
{\cal C}_1(L_\Sigma)=[B/2\pi]\in H^2(\Sigma;{\bb Z})\cong{\bb Z}
\label{chernclass}\end{equation}
which is labelled by the first Chern number
\begin{equation}
c_1(L_\Sigma)=\frac1{2\pi}\int_\Sigma B\in{\bb Z}
\label{chernnum}\end{equation}
When $c_1(L_\Sigma)\neq0$, the gauge field $C|_\Sigma$ is not a function on
$\Sigma$ but rather a section of the non-trivial line bundle
$L_\Sigma\to\Sigma$. Its curvature can be written as $dC|_\Sigma=dC'+2\pi{\cal
C}_1(L_\Sigma)$, where $C'$ is a single-valued one-form on $\Sigma$. The class
of connections $C'$ is the $2g$-dimensional torus $H^1(\Sigma;{\bb
R})/H^1(\Sigma;{\bb Z})\cong{\bb R}^{2g}/{\bb Z}^{2g}\cong T^{2g}$, where the
real cohomology accounts for all canonical locally gauge-equivalent connections
while the integer cohomology accounts for equivalence under large gauge
transformations.

We shall be interested in a particular process which shifts the Chern number
(\ref{chernnum}) and hence maps onto a new line bundle
$\widetilde{L}_\Sigma\to\Sigma$. To see how this can arise, we consider the
Hodge decomposition for the one-form $C|_\Sigma=C_i(x)dx^i$ on $\Sigma$ at
fixed time $t$,
\begin{equation}
C|_{\Sigma}=\frac{4\pi}{k'}\,d\xi+*d\left(
\frac{1}{\nabla^2_\perp}B\right)+4\pi i\,\gamma(t)
\label{hodge}\end{equation}
where
\begin{equation}
k'=k+2
\label{kprime}\end{equation}
Here $\xi$ is some function on $\Sigma$, $\nabla_\perp^2$ is the scalar
Laplace-Beltrami operator with its zero modes removed, and $\gamma$ is a
harmonic one-form on $\Sigma$, $d\gamma=d*\gamma=0$. This harmonic form can be
expanded as
\begin{equation}
\gamma(t)=\sum_{l=1}^{g}\left(\bar{\gamma}^l(t)\omega_l-\gamma^l(t)
\bar{\omega}_l\right)
\end{equation}
where $\omega_l$ are holomorphic harmonic 1-forms which generate
$H^{1,0}(\Sigma)$. They obey the canonical period normalizations
\begin{equation}
\oint_{a^m}\omega_l=\delta_{lm}~~~~~~,~~~~~~\oint_{b^m}\omega_l=\Omega_{lm}
\label{periodnorms}\end{equation}
where $a^l,b^l$, $l=1,\dots,g$, form a basis of canonical homology cycles, i.e.
$a^l\cap b^m=\delta^{lm}$, $a^l\cap a^m=b^l\cap b^m=0$, and
\mbox{$\Omega_{lm}$} is the \mbox{$g\times g$} symmetric period matrix, with
\mbox{Im\,$\Omega>0$}, which is a function of the modular parameters of
$\Sigma$. The metric on the space of holomorphic harmonic 1-forms is
\begin{equation}
G_{lm}\equiv i\int_\Sigma\omega_l\wedge\bar{\omega}_m=2\,\mbox{Im}\,\Omega_{lm}
\end{equation}

Taking the exterior derivative of (\ref{hodge}) we have
\begin{equation}
dC|_\Sigma=d^2\theta+B(x)\,d^2x
\label{dhodge}\end{equation}
where $\theta=(4\pi/k')\xi$ is the pure gauge degree of freedom of $C$ on
$\Sigma$. Since $d$ is nilpotent on $C^\infty(\Sigma)$, for smooth functions
$\theta$ in (\ref{hodge}) it follows that $\int_\Sigma
dC|_\Sigma/2\pi=c_1(L_\Sigma)$ is just the Chern number of the original line
bundle $L_\Sigma$. However, we could just as easily allow for a multi-valued
function $\theta(x)$ on $\Sigma$. The simplest case is when
$\theta(x)=n\theta(z,z_0)$ (with $z\equiv x^1+ix^2$) is the angle function of
the Riemann surface \cite{bergeron} with winding number $n\in{\bb Z}$ around a
fixed point $z_0\in\Sigma$. It is the multi-valued function which is related to
the prime form of $\Sigma$,
\begin{equation}
\frac1{\nabla^2}\,\delta_\Sigma^{(2)}(z-z_0)=-\frac1\pi\log {\cal E}(z,z_0)
\label{primeform}\end{equation}
where
\begin{equation}
{\cal E}(z,z_0)=\frac{\Theta^{(g)}\pmatrix{1/2\cr1/2\cr}
\left(\int_{z_0}^z\omega\Bigm|\Omega\right)}{\sqrt{h(z)h(z_0)}}
{}~~~~~~{\rm with}~~h(z)=\omega^l(z)\frac\partial{\partial
u^l}\Theta^{(g)}\pmatrix{1/2\cr1/2\cr}(u|\Omega)\biggm|_{u=0}
\label{primetheta}\end{equation}
and
\begin{equation}
\Theta^{(g)}\pmatrix{\alpha\cr\beta\cr}(z|\Omega)=\sum_{\{n_l\}\in{\bbs
Z}^g}\exp\left[i\pi(n_l+\alpha_l)\Omega^{lm}(n_m+\alpha_m)+2\pi
i(n_l+\alpha_l)(z^l+\beta^l)\right]
\label{thetadef}\end{equation}
are the holomorphic, doubly semi-periodic Jacobi theta-functions of
$\{z_l\}\in{\bb C}^g$, with $\alpha_l,\beta^l\in[0,1]$. The prime form ${\cal
E}(z,z_0)$ is antisymmetric in $(z,z_0)$ and for $z\sim z_0$ it behaves as
${\cal E}(z,z_0)\sim z-z_0$. Then
\begin{equation}
\theta(z,z_0)={\rm Im}\,\log\left(\frac{{\cal E}(z,z_0)}{{\cal E}(z,z'){\cal
E}(z',z_0)}\right)
\label{anglefndef}\end{equation}
so that
\begin{equation}
d\theta=*d\log{\cal E}
\label{thetaEdiff}\end{equation}
where $z'$ is an arbitrary fixed reference point. The angle function satisfies
the Laplace equation on $\Sigma$ and is not differentiable at the point $z_0$,
\begin{equation}
\nabla^2\theta(z,z_0)=0~~~~~~,~~~~~~d^2\theta(z,z_0)=2\pi\delta_\Sigma^{(2)}
(z-z_0)\,d^2z
\label{thetadiffs}\end{equation}
$\theta(z,z_0)$ is a single-valued function on the universal covering space of
$\Sigma-\{z_0\}$ and along any closed contour ${\cal C}\subset\Sigma$ winding
around the point $z_0$ it has the property $\oint_{\cal C}d\theta=2\pi~{\rm
mod}~2\pi$, i.e. $\theta(z,z_0)\in[0,2\pi)$. The definition of $\theta(z,z_0)$
in terms of the prime form ${\cal E}(z,z_0)$ is just the generalization to an
arbitrary Riemann surface of the angle function on the plane, i.e. the angle of
the line joining $z$ to $z_0$ relative to a fixed reference axis (determined by
$z'$ here).

With this choice of function $\theta$, (\ref{dhodge}) becomes
\begin{equation}
d\widetilde{C}|_\Sigma=\left[2\pi n\delta_\Sigma^{(2)}(z-z_0)+B(z)\right]\,d^2z
\label{multiC}\end{equation}
and consequently
\begin{equation}
\frac1{2\pi}\int_\Sigma d\widetilde{C}|_\Sigma=n+c_1(L_\Sigma)\equiv
c_1(\widetilde{L}_\Sigma)
\label{chernshift}\end{equation}
Thus the effect of making the pure gauge degree of freedom of $C$ a compact
variable is to shift the Chern number $c_1$ and hence the topological class
(\ref{chernclass}) of the line bundle, i.e. it produces a {\it different} line
bundle $L_\Sigma\to\widetilde{L}_\Sigma$. This topology changing process on the
principal fiber bundle of the Chern-Simons gauge theory will be the essence
behind the mirror transformation in the three-dimensional description.

\newsubsection{Monopole-instantons in the Hamiltonian formalism}

The topological consequences of the compactification of the Chern-Simons gauge
theory above are due to the appearence of magnetic monopoles in the theory. The
main idea is that one can achieve the multi-valued shift of the Chern numbers
by considering a {\it smooth} scalar field $\xi$ in the Hodge decomposition
(\ref{hodge}) and making the $U(1)$ gauge group compact. The gauge group now
contains the non-trivial topological information and its effect on the Hilbert
space of the gauge theory is to effectively shift the functions $\xi$ by the
multi-valued angle function on $\Sigma$. For this, we consider the abelian
topologically massive gauge theory \cite{tmgt}
\begin{equation}
S_{\rm TMGT}[C] = \int_{\cal M}\left(-\frac{1}{4e^2}F(C)\wedge\star F(C)+
\frac{k'}{8\pi}C\wedge F(C)+C\wedge\star J\right)
\label{tmgt}\end{equation}
for the $U(1)$ gauge field $C$ coupled to a current $J$ representing the
propagation of a charged particle in the bulk. The kinetic term for $C$
explicitly breaks the topological invariance of the pure gauge theory. It is
included for full generality because radiative corrections by dynamical matter
fields coupled to a Chern-Simons gauge field induce a Maxwell term for it.
Furthermore, its presence allows for the construction of different string
worldsheet actions, including the action for the heterotic string, using the
topological membrane approach to string theory \cite{heterotic}, and it also
enables one to vary the choice of worldsheet complex structure in the induced
conformal field theory on $\Sigma$ via its coupling to the metric of $\cal M$
\cite{tm}. The dimensionful parameter $e^2$ can be thought of as a regulator
such that at the end of calculations one takes the limit $e^2\to\infty$ to
recover the original pure Chern-Simons gauge theory.

Canonical quantization of (\ref{tmgt}) in the Weyl gauge ($C_0=0$) gives the
equal-time commutation relations
\begin{equation}
\left[\Pi_i(x),C_j(y)\right] = i\delta_{ij}\delta_\Sigma^{(2)}(x-y)
\label{ccr}\end{equation}
where
\begin{equation}
\Pi_i=-\frac1{e^2}E_i-\frac{k'}{8\pi}\epsilon_{0ij}C^j
\label{canmom}\end{equation}
is the canonical momentum conjugate to the gauge field $C$, and $E_i=\dot C_i$
is the electric field. The operator
\begin{equation}
{\cal G}=\frac{1}{e^2}\partial_iE^i-\frac{k'}{4\pi}B-J^0
\label{constraint}\end{equation}
generates time-independent local gauge transformations, and the elements of the
local gauge group are the operators
\begin{equation}
U=\exp\left\{-i\int_\Sigma d^2x~\theta(x)\left(\frac{1}{e^2}
\partial_iE^i-\frac{k'}{4\pi}B-J^0\right)\right\}
\label{group}\end{equation}
with $UCU^{-1}=C+d\theta$, i.e.
\begin{equation}
U\,\xi\,U^{-1}=\xi+(k'/4\pi)\theta
\label{xishift}\end{equation}
The physical Hilbert space of the theory contains only those states
$|\Psi\rangle$ which are gauge-invariant, $U|\Psi\rangle=|\Psi\rangle$.

If the gauge group is compact, however, we must be more careful. As discussed
in \cite{monoinst,kaiming}, we must also include in the gauge group the
operators $V$ of the form (\ref{group}) where $\theta(x)$ is a multi-valued
function on $\Sigma$. When $\theta$ is the angle function (\ref{anglefndef}),
we denote the corresponding gauge group element by $V(x_0)$. Then integrating
by parts in (\ref{group}) implies that the physical states must also be fixed
points of the operators
\begin{equation}
V(x_0)=\exp\left\{-i\int_\Sigma
d^2x~\left[\left(\frac1{e^2}E^i+\frac{k'}{4\pi}\epsilon^{0ij}A_j\right)
\epsilon_{0ik}\,\partial^k\log {\cal E}(x,x_0)-\theta(x,x_0)J^0\right]\right\}
\label{vortexop}\end{equation}
In this representation the operators (\ref{vortexop}) commute with non-compact
gauge transformations and also amongst themselves for different points
$x_0\in\Sigma$, so that the invariance condition
$V(x_0)|\Psi\rangle=|\Psi\rangle$ can be imposed simultaneously for all $x_0$
(as required for rotational and translational invariance of the physical
Hilbert space). The identities (\ref{thetadiffs}) along with the commutation
relations (\ref{ccr}) imply that
\begin{equation}
\left[B(x),V^n(x_0)\right]=2\pi n\,\delta_\Sigma^{(2)}(x-x_0)\,V^n(x_0)
\label{vortex}\end{equation}
Thus the operator $V^n(x_0)$ creates a pointlike magnetic vortex at the point
$x_0$ on the worldsheet $\Sigma$ with flux $\Phi=(1/2\pi)\int_\Sigma B=n$,
where $n$ is the monopole number. Since the electric field decays exponentially
at large distances (the photon in (\ref{tmgt}) is massive), from (\ref{flux})
we find that the operator $V^n(x_0)$ also creates electric charge
\begin{equation}
\Delta Q=-nk'/2
\label{viol}\end{equation}
This change of charge/flux is, however, unobservable far from the vortex
because local observables such as the electric field fall off exponentially and
the Aharonov-Bohm phase is unity. Therefore $V(x_0)$ is the operator for a
monopole-instanton \cite{monoinst} which is a dyon that interpolates between
topologically inequivalent vacua of the topologically massive gauge theory. The
magnetic monopoles appear here as topologically non-trivial configurations of
the gauge group when the vacuum is projected onto the gauge-invariant subspace
of the Hilbert space.

The magnetic monopole-instantons give an explicit kinematical realization of
the discrete symmetry of the $N=2$ superconformal minimal model as the magnetic
flux symmetry group of the compact topologically massive gauge theory. The
Bianchi identity
\begin{equation}
dF(C)=0
\label{bianchi}\end{equation}
ensures the existence of a conserved topological current $\star dC$, whose
associated global charge is precisely the flux $\Phi$. It generates the
magnetic flux group of the gauge theory (a subgroup of the gauge group in the
source free $e^2\to\infty$ limit). However, because the monopole-instantons
change the magnetic flux by an integer, only a discrete subgroup of the flux
group, consisting of the operators
\begin{equation}
U_\ell=\e^{2\pi i\ell\Phi}~~~~~~,~~~~~~\ell\in{\bb Z}
\label{fluxsubgp}\end{equation}
remains a symmetry group of the theory. Since $\Phi$ is quantized in units of
$1/k'$, it is only the operators (\ref{fluxsubgp}) for $\ell=0,1,\dots,k'-1$
that are represented non-trivially on this subspace of the physical Hilbert
space. Therefore the magnetic flux symmetry group of compact topologically
massive gauge theory is ${\bb Z}_{k'}$, which is precisely the holomorphic (or
anti-holomorphic) statistical component of the discrete symmetry group ${\bb
Z}_{k'}\times\bar{\bb Z}_{k'}\times{\bb Z}_2\times\bar{\bb Z}_2$ of $M_k$. In
the next section we shall describe the orbifold conformal field theory
$M_k/{\bb Z}_{k'}$ explicitly in the three-dimensional language. This
orbifolding is required since the effects of the magnetic monopoles are
unobservable.

\begin{figure}[htb]
\centerline{\psfig{figure=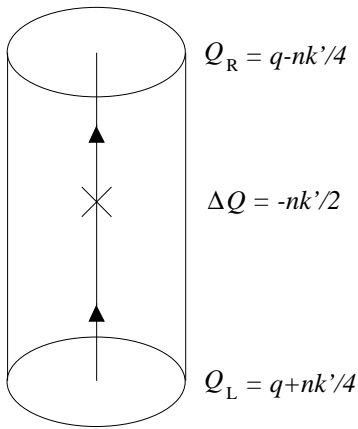,height=0.35\textwidth}}
\small
\caption{A monopole-instanton can change the charge by $nk'/2$ along a Wilson
trajectory.}
\label{monopole}\end{figure}

The presence of monopole-instantons also shifts the spectrum of allowed charges
in the compact theory. To see this, we first write the gauge group generators
in (\ref{group}) in terms of the canonical momenta $\Pi^i$, and in the
functional Schr\"odinger representation where $\Pi^i=i\,\delta/\delta C_i$ it
is easy to see that physical states acquire a non-trivial projective phase
under gauge transformations
\begin{equation}
V(x_0)\Psi[C]=\exp\left\{i\int_\Sigma d^2x~\theta(x,x_0)\left(\frac{k'}{8\pi}B+
J^0\right)\right\}\,\Psi[C+d\theta]
\label{phase}\end{equation}
where $\Psi[C]=\langle C|\Psi\rangle$ with respect to the field basis. The
cocycle of the gauge group in (\ref{phase}) should be single-valued under the
shift $\theta\to\theta+2\pi$. This implies that in the compact theory the
charge $Q=\int_\Sigma d^2x~J^0$ must be quantized according to
\begin{equation}
Q =q+\frac{k'}{8\pi}\int_\Sigma B
\label{quant}\end{equation}
where $q$ is an integer representing the particle winding number around the
monopole-instanton at $x_0\in\Sigma$, i.e. invariance of the quantum field
theory under large gauge transformations. When $k'=0$ we recover the usual
Dirac charge quantization condition of compact quantum electrodynamics. The
extra term in (\ref{quant}) is due to the monopole-instanton background which,
according to (\ref{vortex}), carries $n$ units of magnetic flux. Thus the
spectrum of allowed charge in the compact gauge theory is
\begin{equation}
Q=q+nk'/4
\label{charge}\end{equation}
We can therefore consider the charge non-conservation process illustrated in
fig. \ref{monopole}. A particle of charge \mbox{$Q_{\rm L}=q+nk'/4$} is
inserted on the left boundary and then propagates through the bulk. It then
interacts with the monopole-instanton background which itself carries charge
(\ref{viol}). The effect of the monopole-instanton is to change the charge of
the particle, creating a new state of charge \mbox{$Q_{\rm R}=q-nk'/4$} which
then propagates to the right boundary.

\bigskip

\bigskip

\newsubsection{Functional Schr\"odinger representation}

It is instructive to examine these properties using the functional
Schr\"odinger representation of the Hilbert space. Here we shall use the
approach of \cite{bergeron} for the most part. For this, we consider the limit
$1/e^2\to0$ whereby the full topologically massive gauge theory (\ref{tmgt})
reduces to the pure Chern-Simons theory for $C$ which is an exactly solvable
three-dimensional topological field theory. This Hilbert space corresponds to
the vacuum sector of the topologically massive gauge theory and it coincides
with the moduli space of flat gauge connections on the Riemann surface $\Sigma$
\cite{witten}. We recall from subsection 4.1 that it was precisely this space
which characterized the topology of the Chern-Simons gauge theory. For a
monopole number $n\neq0$, the contributions from the various vacuum sectors are
summed over in the partition function to yield a dynamical representation of
this moduli space.

We decompose the current \mbox{$J$} similarly to (\ref{hodge}) as
\begin{equation}
*J= -d\chi+*d\psi +i\sum_{l=1}^{g}\left(\bar{j}^l
\omega_l-j^l\bar{\omega}_l\right)
\label{current}\end{equation}
The continuity equation $d\star J=(\partial J^0/\partial t)\,d^3x+d*J\wedge
dt=0$ then yields $\psi=-\nabla_\perp^{-2}(\partial J^0/\partial t)$. Using
(\ref{hodge}) we find that the equal-time canonical commutation relations
(\ref{ccr}) in the limit $1/e^2\to0$ can be written as
\begin{equation}
[\xi(z),B(w)] = -i{\cal P}\,\delta_\Sigma^{(2)}(z-w) \ \ \ \ \mbox{or} \ \ \ \
B(z)=i{\cal P}\frac{\delta}{\delta\xi(z)}
\end{equation}
and
\begin{equation}
[\gamma_l,\bar{\gamma}_m]=\frac{1}{4\pi k'}G_{lm}\ \ \ \ \mbox{or}\ \ \ \
\bar{\gamma}_l= -\frac{1}{4\pi k'}G_{lm}\frac{\partial}{\partial\gamma_m}
= -\frac{1}{4\pi k'}\frac{\partial}{\partial\gamma^l}
\end{equation}
where $\cal P$ is the projection operator onto the space orthogonal to the
zero modes of the scalar Laplace-Beltrami operator. The elements of the local
gauge group are the operators
\begin{equation}
U=\exp\left\{i\int_{\Sigma}d^2z~\theta\left(\frac{ik'}{4\pi}{\cal P}
\frac{\delta}{\delta\xi}+J^0\right)\right\}
\label{gaugegroup}\end{equation}

The Hamiltonian operator separates into a local part and a commuting
topological part
\begin{equation}
H=-\int_{\Sigma}C\wedge *J= H_{\rm loc}+H_{\rm top}
\end{equation}
where
\begin{equation}
H_{\rm loc}=\int_{\Sigma}d^2z~\left(i\chi{\cal P}\frac{\delta}{\delta\xi}-
\frac{4\pi}{k'}\xi\frac{\partial J^0}{\partial t}\right)~~~~~~,~~~~~~H_{\rm
top}=i\left(4\pi\bar{j}_l\gamma^l+\frac{1}{k'}j^l
\frac{\partial}{\partial\gamma^l}\right)
\end{equation}
We can therefore use separation of variables and write the wavefunctions as
\begin{equation}
\Psi(\xi,\gamma;t)=\Psi_{\rm loc}(\xi,t)\Psi_{\rm top}(\gamma,t)
\label{wavefull}\end{equation}
The Gauss' law constraint ${\cal G}\approx0$ at $1/e^2\to0$ acts only on the
local part of the wavefunction and is solved by
\begin{equation}
\Psi_{\rm loc}(\xi,t)=\exp\left(\frac{4\pi i}{k'}\int_{\Sigma}d^2z~
\xi(z)J^0(z,t)\right)\Psi_{\rm loc}(t)
\label{localwave}\end{equation}
The solution to the first Schr\"odinger equation $i\partial\Psi_{\rm
loc}(\xi,t)/\partial t=H_{\rm loc}\Psi_{\rm loc}(\xi,t)$ is then given by
\begin{equation}
\Psi_{\rm loc}(t)=\exp\left(\frac{4\pi i}{k'}\int_0^tdt'~\int_{\Sigma}
d^2z~\chi(z,t')J^0(z,t')\right)
\end{equation}
The wavefunctions (\ref{localwave}) carry the usual one-dimensional unitary
representation of the local $U(1)$ gauge group which acts on (\ref{localwave})
by
\begin{equation}
\Psi_{\rm loc}(\xi+(k'/4\pi)\theta,t)=\e^{i\int_\Sigma
d^2z~\theta(z)J^0(z,t)}\,\Psi_{\rm loc}(\xi,t)
\label{locgtwavefn}\end{equation}
The topological part $\Psi_{\rm top}(\gamma,t)$ of the full wavefunction will
be examined in the next section.

When the gauge group is compact, we must also include in it the operators
representing monopole-instantons which can change the charge of a particle
along a Wilson trajectory. To see this explicitly, we write the elements of the
compact gauge group as the exponential of the Gauss' law
\begin{equation}
V(z_0) = \exp\left\{ i\int_{\Sigma}\theta(z,z_0)\left(\frac{k'}{4\pi}
dC+*J^0\right)\right\}
\end{equation}
Substituting in the Hodge decomposition (\ref{hodge}) and integrating by
parts gives
\begin{equation}
V(z_0)=U\exp\left\{-i\int_{\Sigma}\xi\,d^2\theta(z,z_0)\right\}
\end{equation}
where $U$ is given by (\ref{gaugegroup}). Using (\ref{thetadiffs}), it follows
that on the subspace of the Hilbert space of states invariant under the
non-compact part of the gauge group, we must also include the actions of the
operators
\begin{equation}
V^n(z_0)=\exp\left\{-2\pi
in\int_{\Sigma}d^2z~\xi(z)\,\delta_\Sigma^{(2)}(z-z_0)\right\}
\end{equation}
The effect of these operators acting on the wavefunctions (\ref{localwave})
is to shift the charge density by $nk'/2$ at the position $z_0\in\Sigma$ of the
monopole-instanton,
\begin{equation}
V^n(z_0)\Psi_{\rm loc}(\xi,t)=\exp\left\{\frac{4\pi i}{k'}
\int_{\Sigma}d^2z~\xi(z)\left(J^0(z,t)-\frac{nk'}{2}\delta_\Sigma^{(2)}
(z-z_0)\right)\right\}\Psi_{\rm loc}(t)
\end{equation}

This gives an analytic representation of the monopole-instanton charge inducing
process depicted in fig. \ref{monopole}. We start in the holomorphic sector
$\Sigma_{\rm L}$ with an initial wavefunction $\Psi_{\rm L}(\xi)\equiv\Psi_{\rm
loc}(\xi,0)$ of charge $Q$ quantized according to (\ref{charge}). At a certain
time we apply the compact gauge transformation corresponding to the interaction
with the monopole-instanton background localized around the point
$z_0\in\Sigma$, arriving finally in the anti-holomorphic sector $\Sigma_{\rm
R}$ with final wavefunction $\Psi_{\rm R}(\xi)\equiv V^n(z_0)\Psi_{\rm
loc}(\xi,1)$ of charge $Q-nk'/2$. For a point particle moving on $\Sigma$ with
trajectory $z(t)$, we have
\begin{equation}
V^n(z_0)\Psi_{\rm loc}(\xi,t)=\exp\left\{\frac{4\pi
i}{k'}\left(Q\xi(z(t))-\frac{nk'}2\xi(z_0)\right)\right\}\Psi_{\rm loc}(t)
\end{equation}
so that this process is achieved if we take the interaction of the particle
with the monopole-instanton at $z(1)\equiv z_0$.

\newsection{Mirror maps and duality in Chern-Simons theory}

The map between marginal operators ${\cal O}_{(1,1)}\leftrightarrow{\cal
O}_{(1,-1)}$, which is the key to constructing the mirror manifold associated
with the $N=2$ superconformal minimal model, is trivial from the perspective of
the two-dimensional conformal field theory as it is simply the change of sign
of the $U(1)_{\rm R}$ charge associated with each marginal operator. However,
it provides a non-trivial and unexpected map between the corresponding
Calabi-Yau manifolds which have different topologies. We shall now see that the
mirror map
also implies a non-trivial map between Chern-Simons gauge theories of different
topologies. In this way the phenomenon of mirror symmetry, which asserts the
existence of two topologically inequivalent target spaces associated with the
{\it same} conformal field theory, also implies the existence of two
inequivalent topological membranes that induce precisely the same conformal
field theory. We shall also describe the relationship between the mirror map
and a certain topological duality symmetry of the three-dimensional gauge
theory.

\newsubsection{Deformed cohomology rings and the mirror transformation}

We begin by describing the chiral rings of the minimal model $M_k$ as a
deformation of the conformal field theory represented by the coupling of
charged particles to the compact $U(1)_{k+2}$ Chern-Simons gauge theory. As
explained in \cite{us}, the left- and right-moving magnetic fluxes must be
quantized as odd integers. The odd-integer flux quantization condition yields
the fermion-boson transmutation property of the theory, because it produces an
additional factor of $-1$ in the Aharonov-Bohm phases (\ref{2partwave}) which
maps bosons into fermions and vice versa.\footnote{This property will be
especially important in section 6 where we study the quantum Hall effect.} As
described in section 2, the pair of integer-valued fluxes
\begin{equation}
(\Phi_{\rm L},\Phi_{\rm R})=(i,-j)\in{\bb Z}_{d+1}\times{\bb Z}_{d+1}
\label{intfluxes}\end{equation}
labels the Hodge numbers $h^{i,j}$ of the corresponding geometrical
representation. Here $d$ is some integer which, for illustrative purposes, we
identify as a ``dimension" for the time being. With this convention $j<0$ label
$h^{i,d-|j|}$. At the end of this subsection we shall see how to map the
minimal model onto a genuine Calabi-Yau manifold where $d$ will be a true
complex dimension.

For generic $\Phi_{\rm L}\neq\Phi_{\rm R}$ we have corresponding charges
$Q_{\rm L}^{(i,-j)}\neq Q_{\rm R}^{(i,-j)}$ and so we need to incorporate
non-trivial monopole-instanton effects to produce a charge non-conservation
process. The charges $Q_{\rm L}^{(i,-j)}$ are quantized according to
(\ref{charge}) and $Q_{\rm R}^{(i,-j)}=Q_{\rm L}^{(i,-j)}-n^{(i,-j)}k'/2$ for
some monopole numbers $n^{(i,-j)}\in{\bb Z}$. Given the charge-flux
relationship (\ref{flux}), we have explicitly
\begin{equation}
q^{(i,-j)}=-\mbox{$\frac{k'}4$}\,(i-j)~~~~~~,~~~~~~n^{(i,-j)}=-(i+j)
\label{qijnij}\end{equation}
The three-dimensional description of the chiral-chiral ring is now evident.
First note that the horizontal and vertical symmetries (\ref{dualhij}) of the
Dolbeault cohomology ring imply that $(i,j)=(j,i)$ and $(-i,-j)=(i,j)$. It
therefore suffices to consider only positive values of $i$ and $j\in[-i,i]$,
since, as discussed in section 3, the other pairs of fluxes can then be
obtained by parity and time-reversal transformations of the 3-manifold ${\cal
M}=\Sigma\times[0,1]$. We now decompose the complex line bundle
$L_\Sigma\to\Sigma$ described in the previous section into a Whitney sum of
$(\frac d2+1)^2$ line bundles,
\begin{equation}
L_\Sigma=\bigoplus_{i=0}^{d/2}~\bigoplus_{j=-i}^iL^{(i,j)}_\Sigma
\label{Lijdef}\end{equation}
The line bundle $L_\Sigma^{(i,-j)}\to\Sigma$ has Chern number
$c_1(L_\Sigma^{(i,-j)})=n^{(i,-j)}=-(i+j)$, so that
\begin{equation}
c_1(L_\Sigma)=\mbox{$-\frac d{12}(\frac d2+1)(2d+5)$}
\label{totchern}\end{equation}
Note that even though many of the component line bundles in (\ref{Lijdef}) have
the same topological class, it is necessary to incorporate them all to describe
the full cohomology ring. For a given monopole number $n$, with $-d\leq
n\leq0$, it follows from (\ref{hodgeduality}) that the line bundles in the
topological class labelled by $n$ generate the $n$-th Betti number $b^n=\dim
H^n$ via
\begin{equation}
b^n=\sum_{n^{(i,-j)}=n}h^{i,j}
\label{bettihodge}\end{equation}

Corresponding to (\ref{Lijdef}) we have an orthogonal decomposition of the
Chern-Simons gauge field into a sum $C=\sum_{i,j}C^{(i,j)}$, where $C^{(i,j)}$
is a gauge connection on $L^{(i,j)}_\Sigma$. We then introduce $\frac d2+1$
charges $Q^{(i)}=-\frac{k'}2i$, which for each $i$ minimally couples to the
$2i+1$ Chern-Simons gauge fields $C^{(i,j)}$.\footnote{We stress that $C$ is
still regarded here as a {\it single} $U(1)$ gauge field which is expanded into
its components associated with the topological decomposition (\ref{Lijdef}). At
the end of this subsection we will combine several $U(1)_{k+2}$ gauge fields
$C$ to make contact with Calabi-Yau sigma-models.} Thus the relevant part of
$M_k$ which describes the deformed cohomology ring is given in
three-dimensional terms as
\begin{equation}
k'{\cal S}^{(d)}[C]=\sum_{i=0}^{d/2}~\sum_{j=-i}^ik'S_{\rm
CS}^{[U(1)]}[C^{(i,j)}]+\int_{\cal
M}~\sum_{i=0}^{d/2}J^{(-k'i/2)\mu}\sum_{j=-i}^iC_\mu^{(i,j)}
\label{csdcraction}\end{equation}
There are then two types of deformations corresponding to bundles with
vanishing or non-vanishing Chern classes. The deformations representing the
spaces of harmonic $(i,d-i)$-forms and $(i,j)$-forms with $j\neq d-i$ are the
Wilson line operators
\begin{equation}
W^{(i,i)}[C]=\exp\left(iQ^{(i)}\int_{\rm L}^{\rm
R}C_\mu^{(i,i)}(x)~dx^\mu\right)~~~~,~~~~W^{(i,-j)}[C]=\exp\left(iQ^{(i)}
\int_{\rm L}^{\rm
R}\hspace{-0.65cm}\times\hspace{0.3cm}C_\mu^{(i,-j)}(x)~dx^\mu\right)
\label{wilsondeforms}\end{equation}
where the cross on the integral indicates the monopole-instanton transition
which changes the charge by $n^{(i,-j)}k'/2$ along the particle trajectory as
it propagates from $\Sigma_{\rm L}$ to $\Sigma_{\rm R}$. In the first process
in (\ref{wilsondeforms}) the magnetic flux of the particle is conserved along
its motion corresponding to the triviality of the line bundle
$L_\Sigma^{(i,i)}\cong\Sigma\times S^1$. In the second process the motion is
such that the fibers of $L_\Sigma^{(i,-j)}$ are twisted non-trivially. The
finite-dimensional Hilbert space of the three-dimensional quantum field theory
(\ref{csdcraction}) then describes the chiral-chiral ring of $M_k$ which leads
to a deformed cohomology ring. This Hilbert space will be discussed in more
detail in the next subsection.

The mirror theory is obtained by letting $j\to-j$, so that the corresponding
particle winding numbers and monopole numbers are related to (\ref{qijnij}) by
\begin{equation}
\widetilde{q}^{(i,j)}=\mbox{$\frac{k'}4$}\,n^{(i,-j)}~~~~~~,~~~~~~\widetilde
{n}^{(i,j)}=\mbox{$\frac4{k'}$}\,q^{(i,-j)}
\label{mirrorqn}\end{equation}
{}From (\ref{mirrorqn}) we see that the mirror map in the topological membrane
description is a process which interchanges particle winding numbers
(associated with large gauge transformations which wind around the
monopole-instanton) and monopole numbers (associated with the Chern classes,
i.e. elements of $H^2(\Sigma;{\bb Z})\cong{\bb Z}$). It is an order-2
transformation that is not an exact symmetry of the Chern-Simons gauge theory,
because of the factors of $k'/4$ that appear in (\ref{mirrorqn}). We can view
this map as a topology changing transformation of the topological membrane. It
interchanges topological components of the Whitney bundle (\ref{Lijdef}),
because it essentially maps
$W^{(i,-j)}[C]\leftrightarrow\widetilde{W}^{(i,j)}[\widetilde{C}]$ which
corresponds to a non-trivial change of topology among the components
\begin{equation}
L_\Sigma^{(i,-j)}\leftrightarrow\widetilde{L}_\Sigma^{(i,j)}
\label{topcompmap}\end{equation}
This topology change on the line bundle $L_\Sigma\to\Sigma$ is in effect the
one described in the previous section corresponding to a change of the Chern
cohomology classes $c_1(L_\Sigma^{(i,-j)})$. The resulting quantum field theory
associated with the mirror line bundle $\widetilde{L}_\Sigma\to\Sigma$ then
describes the chiral-antichiral ring of $M_k$. From this point of view one sees
why the mirror map is ``almost" a symmetry of the three-dimensional theory, in
that the topology of $L_\Sigma$ is insensitive to the particular Whitney
decomposition (\ref{Lijdef}) (since (\ref{totchern}) depends only on the
dimension $d$) and doesn't see this internal topology change of its components.
Thus spacetime topoology change in the three-dimensional picture is a
non-trivial topology change of the principal fiber bundle of the matter-coupled
Chern-Simons gauge theory (\ref{csdcraction}).

\begin{figure}[htb]
\centerline{\psfig{figure=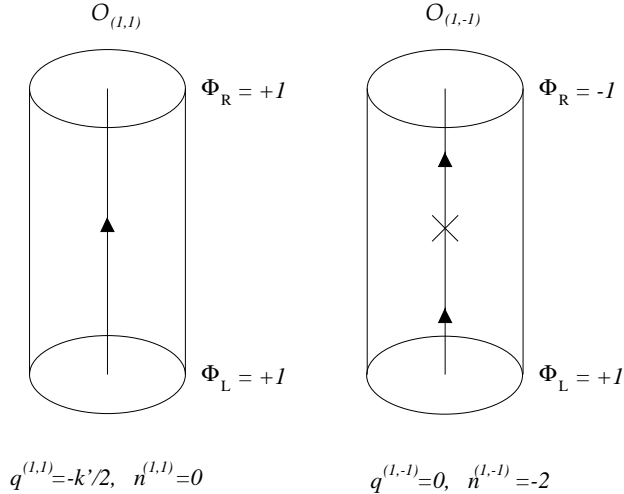,height=0.4\textwidth}}
\small
\caption{Marginal operators ${\cal O}_{(1,1)}$ and ${\cal O}_{(1,-1)}$ as
Wilson
lines in Chern-Simons theory.}
\label{mirror}\end{figure}

As an explicit example, let us examine the deformations by the marginal
operators ${\cal O}_{(1,1)}$ and ${\cal O}_{(1,-1)}$ in this description. The
operator ${\cal O}_{(1,1)}$ has \mbox{${\cal Q}=\bar{\cal Q}=+1$} and can be
represented by the Wilson line
\begin{equation}
{\cal O}_{(1,1)}\sim W^{(1,1)}[C]\end{equation}
describing the propagation of a particle of charge $Q^{(1)}=q^{(1,1)}=-k'/2$
(and $n^{(1,1)}=0$) between left and right boundaries. On the other hand, the
operator ${\cal O}_{(1,-1)}$ has ${\cal Q}=+1$ and $\bar{\cal Q}=-1$ so that
the charge of the particle changes as it propagates between $\Sigma_{\rm L}$
and $\Sigma_{\rm R}$. This is possible if a monopole-instanton with
$n^{(1,-1)}=-2$ (and $q^{(1,-1)}=0$) induces magnetic flux $\Phi_{\rm
L}=-n^{(1,-1)}/2=+1$ on the left boundary and $\Phi_{\rm R}=n^{(1,-1)}/2=-1$ on
the right boundary. This field is thus represented by the Wilson line
\begin{equation}
{\cal O}_{(1,-1)}\sim W^{(1,-1)}[C]
\end{equation}
These processes are depicted in fig. \ref{mirror}.

\bigskip

In terms of the boundary induced conformal field theory, the mirror map is
\begin{equation}
{\cal O}_{(1,1)}\leftrightarrow\widetilde{\cal O}_{(1,-1)}~~~~~~,~~~~~~{\cal
O}_{(1,-1)}\leftrightarrow\widetilde{\cal O}_{(1,1)}
\label{map}\end{equation}
Since both the original conformal field theory and its mirror are induced from
a Chern-Simons theory, the mirror transformation (\ref{map}) gives a map
between gauge theories in (\ref{csdcraction}),
\begin{equation}
\begin{array}{ccc}
W^{(1,1)}[C] & \leftrightarrow & \widetilde{W}^{(1,-1)}[\widetilde{C}] \cr
W^{(1,-1)}[C] & \leftrightarrow & \widetilde{W}^{(1,1)}[\widetilde{C}]
\end{array} \ \ \ \ \mbox{or}\ \ \ \
\begin{array}{ccc}
q^{(1,1)}=-\mbox{$\frac{k'}2$},~n^{(1,1)}=0 & \leftrightarrow &
\widetilde{q}^{(1,-1)}=0,~\widetilde{n}^{(1,-1)}=-2\cr
q^{(1,-1)}=0,~n^{(1,-1)}=-2 & \leftrightarrow &
\widetilde{q}^{(1,1)}=-\mbox{$\frac{k'}2$},~\widetilde{n}^{(1,1)}=0\end{array}
\label{wilson}\end{equation}
The mirror map here is thus a transformation between two dynamical
non-perturbative processes. One is defined on a topologically trivial line
bundle but incorporates non-trivial particle winding modes, while the other
contains no particle windings but has a dynamical mechanism interpolating
between topologically inequivalent vacua of the Chern-Simons gauge theory
(\ref{csdcraction}). In this way, the mirror map relates gauge theories defined
on topologically trivial principal fiber bundles to those on topologically
non-trivial ones. Given that the observables and correlation functions of
Chern-Simons theory are intimately related to the topology of 3-manifolds and
the geometry of Riemann surfaces \cite{witten}, this mirror map on the
topological membrane could serve as a powerful computational tool in these
branches of mathematics.

This is not quite the whole story yet. We have to transport the $N=2$
superconformal field theory defined by the minimal model $M_k$ to the region of
the moduli space where the central charge $c=3d$ coincides with that of the
non-linear sigma-model on a Calabi-Yau manifold of complex dimension $d$. This
is achieved using the Gepner construction \cite{gepner} (see also
\cite{minmirror,witteninst}). Consider the conformal
field theory that is the tensor product of $N=2$ minimal models,
\begin{equation}
M(k_1,\dots,k_s)=\left(\mbox{$\bigotimes_{\ell=1}^s$}\,M_{k_\ell}\right)^{U(1)
^{\rm odd}}
\label{gepnermin}\end{equation}
where the superscript denotes the orbifold projection onto a spectrum of odd
integral total $U(1)$ charges. The central charge of (\ref{gepnermin}) is
$c=\sum_{\ell=1}^sc_{k_\ell}$ so that, according to (\ref{centralk}), the
Chern-Simons coefficients $k_\ell$ in (\ref{gepnermin}) obey the non-linear
constraint
\begin{equation}
\sum_{\ell=1}^s\frac{3k_\ell}{k_\ell+2}=3d
\label{kellconstr}\end{equation}
The conformal field theory (\ref{gepnermin}) is equivalent, by the appropriate
transport via a marginal deformation in the moduli space, to the non-linear
sigma-model on the Calabi-Yau space $M$ defined by the zero locus
\begin{equation}
z_1^{k_1+2}+\dots+z_s^{k_s+2}=0
\label{calabiyaumin}\end{equation}
which is considered as an equation in $W{\bb C}P^{s-1}(\frac
D{k_1+2},\dots,\frac D{k_s+2})$, the weighted complex projective space, where
$D$ is the least common divisor of the $k_\ell'$. The mirror of
(\ref{gepnermin}), obtained by changing the sign of all $U(1)_{\rm R}$ charge
eigenvalues, is the orbifold $M/G$, where $G$ is the subgroup of
$\prod_{\ell=1}^s{\bb Z}_{k_\ell'}$ which acts on $M$ by
\begin{equation}
(z_1,\dots,z_s)~\longrightarrow~(\e^{2\pi in_1/{\cal Q}_1}\,z_1,\dots,\e^{2\pi
in_s/{\cal Q}_s}\,z_s)
\label{GactionCY}\end{equation}
where $n_1,\dots,n_s$ are arbitrary integers such that
$\sum_{\ell=1}^sn_\ell/{\cal Q}_\ell$ is an integer. This latter condition
enforces the $U(1)$ projection in (\ref{gepnermin}) and geometrically it is the
condition of preserving the holomorphic $d$-form of $M$.

For the three-dimensional description, we consider $s$ independent copies of
the matter-coupled minimal model action $\sum_{\ell=1}^s{\cal
I}_{k_\ell}^{(j_\ell,m_\ell)}[A_\ell,B_\ell,C_\ell]$ described in section 3,
where the action for the $C$-field part of the coset is
$\sum_{\ell=1}^sk_\ell'{\cal S}^{(d_\ell)}[C_\ell]$ as described above. The
action of the discrete group $G$ on $M$ is encoded through the fractional
statistics acquired by the external charged particles, or equivalently by the
magnetic flux symmetry group of $\prod_{\ell=1}^sU(1)_{k_\ell'}$. This then
relates, via the mirror map, the topology changing process described above on
the Whitney product bundles
$\bigotimes_{\ell=1}^sL_\Sigma^{[\ell]}\leftrightarrow\bigotimes_{\ell=1}^s
\widetilde{L}_\Sigma^{[\ell]}$ to the spacetime topology change
$M\leftrightarrow\widetilde{M}=M/G$ in terms of Calabi-Yau spaces. In this way
we obtain an intriguing relationship between topologically inequivalent
principal fiber bundles over a 3-manifold and inequivalent Calabi-Yau
manifolds, both leading to identical physical models. In both of these
``spacetime" descriptions of the same conformal field theory, the mirror map is
{\it only} a symmetry of the model at the level of the worldsheet theory.

\newsubsection{Orbifold constructions and topological duality}

We now turn once again to the Schr\"odinger representation of the gauge theory.
 This will enable an explicit construction of the orbifold
$\widetilde{M_k}=M_k/{\bb Z}_{k'}$, and will show that mirror symmetry is
related to a `topological duality' property of the matter-coupled Chern-Simons
gauge theory. We consider the propagation of a single particle of charge $Q$
and trajectory $z(t)$ on the Riemann surface $\Sigma$ coupled to the
Chern-Simons gauge field $C$, and we focus on the topological part of the
wavefunctional of subsection 4.3. It satisfies the Schr\"odinger wave equation
$i\partial\Psi_{\rm top}(\gamma,t)/\partial t=H_{\rm top}\Psi_{\rm
top}(\gamma,t)$ which is solved by
\begin{eqnarray}
\Psi_{\rm top}(\gamma,t)&=&\exp\left(4\pi\gamma^l\int_0^tdt'~\bar{j}_l(t')
+\frac{4\pi}{k'}\int_0^tdt'~j_l(t')\int_0^{t'}dt''~\bar{j}^l(t'')\right)
\nonumber\\&&\hspace{2cm}\times~\widehat{\Psi}_{\rm
top}\left(\gamma^l+\frac{1}{k'}\int_0^tdt'~j^l(t')\right)
\end{eqnarray}
In the absence of sources the holomorphic wavefunctions $\widehat{\Psi}_{\rm
top}(\gamma)$ must incorporate the invariance of the gauge theory under the
large gauge transformations
\begin{equation}
\gamma^m\to\gamma^m+s^m+\Omega^{ml}t_l
\label{large}\end{equation}
where $s^m$ and $t_m$ are integers representing the winding numbers of the
gauge field around the canonical homology cycles of $\Sigma$. The invariant
wavefunctions are combinations of the Jacobi theta-functions (\ref{thetadef})
\cite{bergeron}
\begin{equation}
\widehat\Psi_{\rm
top}^{(r)}\pmatrix{\alpha\cr\beta\cr}(\gamma|\Omega)=\e^{-2\pi
k'\gamma^m\gamma_m}\,\Theta^{(g)}\pmatrix{(\alpha+k'+r)/k'\cr\beta
\cr}(k'\gamma|k'\Omega)
\end{equation}
where $r_l=1,2,\ldots k'$, and $\alpha_l,\beta^l\in[0,1]$.

For a point charge $Q$, the current is $J^0(z,t)=Q\delta_\Sigma^{(2)}(z-z(t))$,
$J^z(z,t)=\frac12Q\dot{\bar z}(t)\delta_\Sigma^{(2)}(z-z(t))$. The
decomposition (\ref{current}) can then be solved for the local components
$\chi$ and $\psi$ in terms of the angle function of $\Sigma$, while the
harmonic components are given by
\begin{equation}
j^l(t)=Q\,\dot z(t)\,\omega^l(z(t))
\label{topcurrents}\end{equation}
The Hilbert space has finite dimension $(k')^g$ and is spanned by the full set
of wavefunctions (\ref{wavefull}) which are given by \cite{bergeron}
\begin{eqnarray}
\Psi_r^{(Q)}\pmatrix{\alpha \cr \beta
\cr}(\xi,\gamma;t|\Omega) &=& \exp\left\{-2\pi k'\gamma^m\gamma_m+4\pi\gamma^m
\int_0^tdt'~\left[\bar{j}_m(t')-j_m(t')\right]\right.\nonumber\\&&\hspace{1cm}
+\,\frac{4\pi i}{k'}\left[Q\xi(z(t))-\frac{nk'}2\xi(z_0)\right]+\frac{iQ^2}{k'}
\left[\theta(t)-\theta(0)\right]\nonumber\\&&\hspace{1cm}\left.
+\,\frac{\pi}{2k'}\int_0^tdt'~\left[j_m(t')-\bar{j}_m(t')\right]
\int_0^{t'}dt''~\left[j^m(t'')-\bar{j}^m(t'')\right]\right\}\nonumber\\
&& \hspace{1cm}\times\,\Theta^{(g)}\pmatrix{(\alpha+k'+r)/k'\cr\beta
\cr}\left(k'\gamma^m+\int_0^tdt'~j^m(t')\biggm|k'\Omega\right)\nonumber\\&&
\label{wavefntotal}\end{eqnarray}
where
\begin{equation}
\theta(t)={\rm Im}\left[\log\left(\frac{f(t)}{{\cal
E}(z(t),z')^2}\right)+2\int_{z'}^{z(0)}\omega^l
\int_{z(0)}^{z(t)}\left(\omega_l+\bar\omega_l\right)\right]+\pi
\label{selflink}\end{equation}
is the ``angle function" of the curve $z(t)$ on $\Sigma$, with
$f(t)\sim\lim_{\epsilon\to0}{\cal E}(z(t),z(t)+\epsilon)$ a framing of the
trajectory $z(t)$. Note that in (\ref{wavefntotal}) we have also incorporated
the invariance of the physical states under compact gauge transformations as
prescribed in the previous section.

Considering the assembly of point charges $Q^{(i)}$ described in the previous
subsection coupled to the independent gauge fields $C^{(i,j)}$, and
incorporating the rest of the Hodge diamond according to the symmetries
(\ref{dualhij}), the direct sum of all resulting Hilbert spaces forms a vector
space ${\cal H}(M_k)$ which is the three-dimensional version of the deformed
cohomology ring with multiplication of the corresponding wavefunctionals
(\ref{wavefntotal}). The dimension of this ring is
\begin{equation}
\dim{\cal H}(M_k)=(d+2)^2(k+2)^g\equiv\sum_{n=-d}^db^n
\label{dimring}\end{equation}
For a given geometrical model, the coefficients on the left-hand side of
(\ref{dimring}) can be fixed to coincide with the topological dimension on the
right-hand side.

To describe the orbifold projection by ${\bb Z}_{k'}$, we use the fact that
quantities which are multi-valued due to their windings around the location
$z_0\in\Sigma$ of the monopole-instanton can be considered as single-valued
functions on the universal cover of the punctured Riemann surface
$\Sigma-\{z_0\}$. This introduces an extra homology cycle which winds around
the puncture at $z_0$. Correspondingly, this induces extra (real) harmonic
components $\gamma^0$ and $j^0$ into the Hodge decompositions of the gauge
field and the particle current. The effects of winding around the location of
the monopole-instanton are then given by large gauge transformations analogous
to (\ref{large}) with $t_l=0$. When the gauge field configurations wind $s^0$
times around the puncture $z_0$ we find
\begin{equation}
\Psi_r^{(Q)}
\pmatrix{\alpha\cr\beta\cr}(\xi,\gamma^0+s^0,\gamma^m;t|\Omega)=\exp\left(-4\pi
ik's^0-2\pi i\alpha_0s^0\right)
\Psi_r^{(Q)}\pmatrix{\alpha\cr\beta\cr}(\xi,\gamma^0,\gamma^m;t|\Omega)
\label{wave1}\end{equation}
Combining this with the local and global $U(1)$ gauge invariance, we see that
the wavefunctions (\ref{wavefntotal}) augmented by $\gamma^0$ carry a
representation of the magnetic flux symmetry group ${\bb Z}_{k'}$ in such a way
that on the whole they represent gauge-invariance with respect to the
semi-direct product gauge group\footnote{Note that the $\alpha$'s and $\beta$'s
are all free parameters and can be set to zero for a qualitative discussion of
the wavefunction transformation properties. They can be fixed by requiring
worldsheet modular invariance of the physical states \cite{bergeron}.}
\begin{equation}
{\cal U}_{\rm orb}=U(1)\otimes_{\rm S}{\bb Z}_{k+2}
\label{gorb}\end{equation}
with the magnetic flux symmetry regarded as a discrete automorphism group of
the $U(1)$ gauge group. The Chern-Simons theory with gauge group (\ref{gorb})
describes the orbifold current algebra $U(1)_{k+2}/{\bb Z}_{k+2}$
\cite{mooreET,tm,emss}, which in the present case is enough to yield the
three-dimensional description of the orbifold conformal field theory
$\widetilde{M_k}=M_k/{\bb Z}_{k'}$. This shows how the compact Chern-Simons
gauge theory naturally constructs the required mirror of the $N=2$ minimal
model.

We can also consider topologically non-trivial motions of the particle around
$z_0$ in a time span $t$. In this case the current changes according to
\begin{equation}
\mbox{$\int_0^tdt'~j^0(t')\to\int_0^tdt'~j^0(t')+s^0$}
\end{equation}
with $s^0$ the winding number of the particle. Then the wavefunctions transform
as
\begin{eqnarray}
\Psi_r^{(Q)}\pmatrix{\alpha\cr\beta\cr}(\xi,\gamma;t|\Omega)&=&
\exp\left(\frac{16\pi i}{k'}r_0s^0+\frac{4\pi
iQ}{k'}\left(\alpha_0-\alpha^{(0)}\right)s^0+
\frac{4iQ^2}{k'}\left[f(t)-f(0)\right]
\right)\nonumber\\&&\hspace{2cm}\times~
\Psi_r^{(Q)}\pmatrix{\alpha\cr\beta\cr}(\xi,\gamma;0|\Omega)
\label{wave2}\end{eqnarray}
with $\alpha^{(0)}=Q\int_{z'}^{z(0)}\omega^0$. Comparing the transformations
(\ref{wave1}) and (\ref{wave2}) we see that there is a duality between large
gauge transformations and the motion of particles around the puncture $z_0$,
whereby we exchange $k'\leftrightarrow 4/k'$.

The above topological duality map between the large gauge transformations and
the particle windings around the monopole-instanton is similar to the
Chern-Simons mirror map (\ref{mirrorqn}). A particle of charge $Q=q+nk'/4$
winds $q$ times around the puncture $z_0\in\Sigma$ and invariance under large
gauge transformations around $z_0$ gives the quantization of the monopole
number $n$. Topological duality then exchanges $q\leftrightarrow n$ and
$k'\leftrightarrow 4/k'$ which, modulo the transformation of $k'$, is the
mirror map (\ref{mirrorqn}). The transformation properties of the Chern-Simons
wavefunctions above make explicit the realization of the mirror map from an
anyonic symmetry. However, the mirror map is not an exact duality symmetry of
the three-dimensional theory except at the very special point $k'=4$ of the
moduli space where it corresponds to an exact interchange of particle winding
and monopole numbers. In the topological membrane approach to string theory,
where the Chern-Simons coefficient is related to the radius $R$ of the target
space compactification by $k'=4R^2/\alpha'$ \cite{kogan2,kaiming} with
$1/2\pi\alpha'$ the string tension, the point $k'=4$ corresponds to the
self-dual point of the $T$-duality transformation $R\leftrightarrow\alpha'/R$.
This is precisely the three-dimensional realization of the well-known fact (see
for example \cite{aspinwall}) that, for some points in the moduli space, mirror
symmetry is equivalent to $T$-duality. However, this is a very special case and
in general there is no relationship between mirror symmetry and $T$-duality in
the three-dimensional picture.

Note the appearence of an extra self-linking term in (\ref{wave2}) (the last
term in the exponential) as compared to (\ref{wave1}). As we shall see in the
next subsection, this term essentially incorporates the effects of the
monopole-instanton transition and is crucial to the duality properties in the
following sense. When $\Sigma=S^2$ is the Riemann sphere (where there are no
homology cycles and the prime form is ${\cal E}(z,z_0)=z-z_0$ with
$z'=\infty$), the topological duality $q\leftrightarrow n$ can be thought of as
interchanging the cohomology groups $H^1(S^2-\{z_0\};{\bb Z})\cong{\bb Z}$ and
$H^2(S^2-\{z_0\};{\bb Z})\cong{\bb Z}$. This isomorphism is accomplished by
application of the {\it three-dimensional} Hodge duality operator $\star$ on
${\cal M}=\Sigma\times[0,1]$ (signalling again the relation to a $T$-duality).
For genus $g>0$ this is not the case, because $H^1(\Sigma-\{z_0\};{\bb
Z})\cong{\bb Z}\oplus{\bb Z}^{2g}$ and $H^2(\Sigma-\{z_0\};{\bb Z})\cong{\bb
Z}$ are no longer isomorphic. However, we will see below that the self-linking
term in (\ref{wave2}) depends on the genus $g$ of $\Sigma$ in such a way that
it effectively ``absorbs" this residual cohomology of the topological duality
transformation.

The duality properties discussed in this section can be seen more generally and
explicitly on a generic 3-manifold $\cal M$ without boundary from a simple path
integral approach. For this, we consider the generating functional
\begin{equation}
Z[J]=\int DC~D\widetilde{C}~D\lambda~\exp\left\{i\int_{\cal
M}\left(\frac{k'}{8\pi}C\wedge dC+\frac1{8\pi}\lambda\wedge
\left(k'dC-d\widetilde{C}\right)+\frac{1}{k'}\widetilde{C}\wedge\star
J\right)\right\}
\label{pathduality}\end{equation}
defined as a path integral over three gauge fields $C$, $\widetilde{C}$ and
$\lambda$, where $J$ is an external source current. Integrating first over
$\widetilde{C}$ in (\ref{pathduality}) gives the constraint
$d\lambda=(8\pi/k')\star J$. Integrating over $\lambda$ enforces this
constraint, giving the partition function for the matter-coupled Chern-Simons
gauge field $C$ with statistics parameter $k'$,
\begin{equation}
Z[J]={\cal Z}[J]=\int DC~\exp\left\{i\int_{\cal M}\left(\frac{k'}{8\pi}C\wedge
dC+C\wedge\star J\right)\right\}
\label{pathC}\end{equation}
However, if we instead integrate first over $\lambda$ we obtain the constraint
equation $k'dC=d\widetilde{C}$ which is solved by
\begin{equation}
C=\mbox{$\frac1{k'}$}\,\left(\widetilde{C}+\gamma\right)
\label{dualfieldrel}\end{equation}
where $\gamma$ is a harmonic one-form on ${\cal M}$. Integrating over $C$ we
thus obtain another matter-coupled Chern-Simons theory
\begin{equation}
Z[J]=\widetilde{\cal Z}[\widetilde{J}]=\int
D\widetilde{C}~\exp\left\{i\int_{\cal M}\left(\frac1{8\pi
k'}\widetilde{C}\wedge
d\widetilde{C}+\widetilde{C}\wedge\star\widetilde{J}\right)\right\}
\label{pathdualC}\end{equation}
where $\widetilde{J}=\frac1{k'}J$ is the dual source current. The equivalence
between the two representations (\ref{pathC}) and (\ref{pathdualC}) shows more
precisely the effect of the mirror map on the matter-coupled gauge theory.
Moreover, the relationship between the Chern-Simons field $C$ and its dual
$\widetilde{C}$ in (\ref{dualfieldrel}) shows exactly how the topological
duality corresponds to a change in topology of the gauge theory (by adding a
harmonic component $\gamma$).

\newsubsection{Duality in topologically massive gauge theory}

It is possible to see the effects of the monopole-instanton transition from a
purely three-dimensional perspective. For this, we consider the duality
properties of the full topologically massive gauge theory (\ref{tmgt}).
Although the ground state of this theory has rather simple duality
transformation laws, things are more complicated in the regulated theory for
finite $e^2$. The duality properties of this theory were originally discussed
in \cite{procatmgt} and elucidated on in \cite{kogan2}. Using a path integral
approach in the spirit of the previous subsection, it is possible to prove the
equivalence of the two generating functionals
\begin{equation}
{\cal Z}_M[J]=\widetilde{\cal Z}_M[J]~\e^{i{\cal J}[{\cal C}]}
\label{tmgtduality}\end{equation}
where
\begin{equation}
{\cal Z}_M[J]=\int DC~\exp\left\{i\int_{\cal
M}\left(-\frac{1}{4e^2}F(C)\wedge\star F(C)+\frac{k'}{8\pi}C\wedge
F(C)+C\wedge\star J\right)\right\}
\label{tmgtgenfn}\end{equation}
\begin{equation}
\widetilde{\cal Z}_M[J]=\int D\widetilde{C}~\exp\left\{-i\int_{\cal
M}\left(2e^2\left(\frac{k'}{8\pi}\right)^2\widetilde{C}\wedge\star\widetilde{C}
+\frac{k'}{8\pi}\widetilde{C}\wedge F(\widetilde{C})+\widetilde{C}\wedge\star
J\right)\right\}
\label{proca}\end{equation}
are the generating functionals for the topologically massive and
Chern-Simons-Proca gauge theories, and
\begin{equation}
\e^{i{\cal J}[{\cal C}]}=\exp\left\{-\frac{2\pi i}{k'}\int_{{\cal
M}(x)}\int_{{\cal M}(y)}J(x)\wedge\frac d\Box(x-y)\,J(y)\right\}
\label{selflink3d}\end{equation}
is a non-local phase factor, which, when $J$ is the current for a particle with
closed worldline $\cal C$ and charge $Q$, gives the topological self-linking
number
\begin{equation}
{\cal J}[{\cal C}]=\frac{4\pi Q^2}{k'}\oint_{{\cal
C}(x)}\int\!\!\!\int_{\Sigma({\cal C}(y))}\delta_{\cal M}^{(3)}(x-y)
\label{selflinkC}\end{equation}
of the loop $\cal C$ in the 3-manifold $\cal M$ (the double integral in
(\ref{selflinkC}) is taken with the DeRham current of $\cal C$ over a surface
$\Sigma({\cal C})$ spanned by the loop).

The ``dual" of the topologically massive gauge theory (\ref{tmgt}) is therefore
the Chern-Simons theory plus an additional Proca mass term for the gauge field,
which has topological mass $k'e^2/4\pi$. This is true modulo the non-local
self-linking contribution (\ref{selflink3d}), which we recall also appeared in
the Chern-Simons wavefunctions upon application of the mirror map (compare
(\ref{wave2}) and (\ref{wave1})). Here the self-linking contribution is crucial
to the global properties of this duality map. As discussed in \cite{kogan2}, a
charged particle interacting with the Chern-Simons-Proca field carries {\it no}
induced magnetic flux, so that the linking term (\ref{selflink3d}) (with
linking number 1) alters this property so as to yield the charge-flux
relationship (\ref{flux}) of the topologically massive gauge theory.

The integral (\ref{selflinkC}) diverges for $x\sim y$ and must be regularized.
For this, we define a framing of the loop $\cal C$ in terms of a unit vector
$\vec n(t)$ normal to it, where $t\in[0,1]$ is the parameter of the loop. We
then construct a second loop ${\cal C}'$ by deforming $\cal C$ infinitesimally
along this framing, and compute (\ref{selflinkC}) as the linking number of
${\cal C}'$ and $\cal C$. This yields the torsion of the curve $\cal C$
\cite{anyon,polyakov1988},
\begin{equation}
{\cal J}[{\cal C}]=\frac{4Q^2}{k'}\oint_{\cal C}d\vec x\cdot\left(\vec
n\times\dot{\vec n}\right)
\label{torsionC}\end{equation}
Thus the addition of the non-local phase factor to the dual theory amounts to
the addition of the phase
\begin{equation}
{\cal J}[{\cal C}]=\frac{4Q^2}{k'}\int_0^1dt~\dot\theta(t)
\label{Jtheta}\end{equation}
where $\dot\theta=\dot{\vec x}\cdot(\vec n\times\dot{\vec n})$ is the normal
connection (or torsion form) of the curve $\cal C$. The phase factor
(\ref{torsionC}) appears in the form of the charged particle propagating along
a Wilson loop with non-zero point-like flux. In fact, it can be shown
\cite{polyakov1988} that $\dot\theta(t)$ is essentially a Dirac potential on
the sphere $(\dot{\vec x})^2=1$ in $\cal M$ with a magnetic monopole located at
its center. This shows that the transformation between the topologically
massive gauge theory (\ref{tmgt}) and its dual is also due to some
non-perturbative, topological process induced by the interactions between the
charged particles and a magnetic monopole in the 3-manifold $\cal M$, and
furthermore that this duality is achieved by a deformation in the
three-dimensional language as described in subsection 5.1 above.

On a product manifold ${\cal M}=\Sigma\times{\bb R}$, the expression
(\ref{Jtheta}) coincides with the angle function term in the wavefunction
(\ref{wavefntotal}). Thus the monopole-instanton induced process in the
topologically massive gauge theory is identical to that for its ground state in
the canonical formalism in which the self-linking numbers of the particle
trajectories (see (\ref{wave2})) play a crucial role. Since the framing vectors
of $\cal C$ define a basis in the tangent space, $\dot\theta$ is actually a
connection on the tangent bundle of the Riemann surface $\Sigma$. As the
particle encircles the monopole-instanton at $z_0\in\Sigma$ once, using the
Gauss-Bonnet theorem we find the self-linking contribution
\begin{equation}
{\cal J}[{\cal C}]=\mbox{$\frac{16\pi Q^2}{k'}\,(1-g)$}
\label{selflinkg}\end{equation}
in the transformation law (\ref{wave2}). The magnetic monopoles therefore lead
to highly non-trivial dynamical effects in the vacuum sector of the
topologically massive gauge theory. In addition, by taking the pure
Chern-Simons limit $e^2\to\infty$, the discussion of this subsection
illustrates the origin of the mirror map as a particle winding process, as well
as the appearence of the magnetic flux symmetry from the induced phases
(\ref{selflinkg}) in the wavefunctions (\ref{wave2}), around a
monopole-instanton from a purely three-dimensional perspective.

\newsection{Mirror symmetry in quantum Hall systems}

In this section we shall present a concrete application of the more formal
results that we have developed thus far in this paper to the quantum Hall
effect, for which Chern-Simons gauge theory provides an effective mean field
description \cite{qhecs}. We will see that the mirror map for the class of
toroidal target spaces defined by the linear sigma-model implies an intriguing
transformation between various quantum Hall systems, and, in certain cases, a
non-trivial map between quantum Hall filling fractions. The toroidal
compactifications constitute the simplest examples of Calabi-Yau spaces in
which the mirror transformation can be represented in a completely transparent
form.

\newsubsection{Mirror maps in linear sigma-models}

The mirror map for quantum Hall systems originates from the simplest example of
mirror symmetry, namely the two-dimensional linear sigma-model. We consider a
$U(1)^p$ Chern-Simons gauge theory with fields $A^I$, $I=1,\dots,p$, that have
pure gauge degrees of freedom $\theta^I$. The bulk action is
\begin{equation}
K\cdot S_{\rm CS}^{[U(1)^p]}[A]=\sum_{I,J}\frac{K_{IJ}}{8\pi}\int_{\cal
M}A^I\wedge dA^J
\label{csactionp}\end{equation}
where now $K_{IJ}$ is some real-valued matrix. This action induces on the
boundary $\Sigma=\partial{\cal M}$ the $c=p$ conformal field theory
\cite{kogan2,kaiming}
\begin{equation}
S_{XY}[\theta]=\frac1{8\pi}\int_\Sigma
d^2z~K_{IJ}\partial_z\theta^I\partial_{\bar z}\theta^J
\label{XYaction}\end{equation}
{}From a string theoretic point of view, the matrix $K_{IJ}$ is related to the
background matrix of the $p$-dimensional target space by
\begin{equation}
K_{IJ}=4(g_{IJ}+\beta_{IJ})/\alpha'
\label{backgroundmatrix}\end{equation}
where $g_{IJ}$ is the metric tensor and $\beta_{IJ}$ the torsion form of the
spacetime. The compactification $\theta^I\in[0,2\pi)$ of the fields of
(\ref{XYaction}) yields the $XY$ model which is well-known to possess
non-trivial duality properties \cite{kogan2,sath}.

Henceforth we assume that we are dealing with a {\it compact} $U(1)^p$
Chern-Simons gauge theory, or equivalently that the target space of the induced
conformal field theory (\ref{XYaction}) is a flat $p$-torus $T^p\cong{\bb
R}^p/2\pi\Gamma$, where $\Gamma$ is a compactification lattice of rank $p$. The
notion of a mirror symmetry on such a string background is especially
well-understood and can be given explicitly in terms of a matrix transformation
law for $K_{IJ}$ \cite{giveon}
\begin{equation}
K\to\widetilde{K}\equiv\left[(I_p-E_p)K+\mbox{$\frac4{\alpha'}$}\,E_p\right]
\left[\mbox{$\frac{\alpha'}4$}\,E_pK+(I_p-E_p)\right]^{-1}
\label{mirrormap}\end{equation}
where $I_p$ is the $p\times p$ identity matrix and
$(E_p)_{IJ}=\delta_{Ip}\delta_{Jp}$ is the step operator in the $p$-th
direction of $T^p$. The mirror transformation (\ref{mirrormap}) of the
Chern-Simons coefficient matrix $K$ is actually one of the $p$ factorized
duality maps of compactified bosonic string theory. It can be viewed as a
$T$-duality transformation along the $p$-th direction of $T^p$ by choosing a
particular compactification lattice $\Gamma$ that splits the $p$-torus into the
Cartesian product of a circle of radius $R_p$ and a $(p-1)$-dimensional
background $T^{p-1}$. Then the transformation (\ref{mirrormap}) maps
$R_p\to\alpha'/R_p$ leaving $T^{p-1}$ unchanged \cite{giveon}. However, we
stress again that this equivalence with $T$-duality only occurs at a special
point of the moduli space of toroidal compactifications.

For the three-dimensional description of this mirror map, we couple the allowed
charges
\begin{equation}
Q_I=q_I-\mbox{$\frac12\sum_J$}\,K_{IJ}\,n^J
\label{chargesigma}\end{equation}
to the $I$-th Chern-Simons gauge field $A^I$, whose associated complex line
bundle $L^{(I)}_\Sigma\to\Sigma$ has monopole number $n^I$. Then the mirror map
is precisely the process (\ref{mirrorqn}) which exchanges the particle winding
numbers $q_p$ with the monopole numbers $n^p$ in the $p$-th component of
$U(1)^p$, leaving all of the other $p-1$ components unchanged. Note that the
factorized duality map (\ref{mirrormap}) is defined for {\it any} $p$, even
when $p$ is odd (and hence for compactifications which are not complex
manifolds).

When $p$ is even, as explained in section 2 the mirror map is implemented in
the linear sigma-model by considering an appropriate K\"ahler deformation,
leaving the shape (i.e. the angles between the homology cycles) of $T^p$ fixed
but changing its volume, and also a complex structure deformation, which leaves
the volume fixed but changes the shape \cite{greene}. It exchanges the complex
and K\"ahler structures of $T^p$. To see this, consider the case of the 2-torus
$T^2$. The metric and torsion form can then be written explicitly as
\begin{equation}
[g_{IJ}]=\frac{R_2}{\mbox{Im}\,\tau}\pmatrix{1 & \mbox{Re}\,\tau \cr
\mbox{Re}\,\tau & |\tau|^2}~~~~~~,~~~~~~
[\beta_{IJ}]=\pmatrix{0 & R_1 \cr -R_1 & 0}
\end{equation}
where $R_1\in{\bb R}$, $R_2\in{\bb R}^+$, and $\tau\in{\bb C}^+$ specifies a
complex structure on $T^2$. Written in terms of complex coordinates $z=x^1+\tau
x^2$ we have
\begin{equation}
g=\frac{1}{2}\,\frac{R_2}{\mbox{Im}\,\tau}\,dz \otimes
d\bar{z}~~~~~~,~~~~~~\beta=\frac{i}{2}\,\frac{R_1}{\mbox{Im}\,\tau}\,dz\wedge
d\bar{z}
\end{equation}
The K\"ahler 2-form on $T^2$ is
\begin{equation}
\omega\equiv ig_{z\bar{z}}\,dz\wedge d\bar{z}=\frac{i}{2}\,\frac{R_2}
{\mbox{Im}\,\tau}\,dz\wedge\,d\bar{z}
\end{equation}
with $d\omega=0$. The moduli space of $T^2$ is described by two complex numbers
-- the modular parameter $\tau$ which specifies the complex structure on $T^2$
and also the parameter
\begin{equation}
\sigma=R_1+iR_2=\int_{T^2}(\beta+i\omega)
\end{equation}
which describes the (deformed) K\"ahler structure on $T^2$. Mirror symmetry
interchanges the complex and K\"ahler structures, $\tau\leftrightarrow\sigma$,
or equivalently
\begin{equation}
\mbox{Re}\,\tau\leftrightarrow R_1~~~~~~,~~~~~~
\mbox{Im}\,\tau \leftrightarrow R_2
\end{equation}
In the corresponding Chern-Simons gauge theory, the coefficient matrix
(\ref{backgroundmatrix}) can be written in terms of geometrical parameters as
(here we set $\alpha'=1$ for simplicity)
\begin{equation}
K=\frac{4R_2}{\mbox{Im}\,\tau}\pmatrix{1 & \mbox{Re}\,\tau
+\mbox{$\frac{R_1}{R_2}$}\,\mbox{Im}\,\tau \cr\mbox{Re}\,\tau -
\mbox{$\frac{R_1}{R_2}$}\,\mbox{Im}\,\tau &|\tau|^2}
\end{equation}
In the mirror theory, this matrix becomes
\begin{equation}
\widetilde{K}=\frac{4\,\mbox{Im}\,\tau}{R_2}\pmatrix{1 & R_1 +
\frac{\mbox{Re}\,\tau}{\mbox{Im}\,\tau}\,R_2 \cr
R_1 -\frac{\mbox{Re}\,\tau}{\mbox{Im}\,\tau}\,R_2 & |\sigma|^2}
\label{mirrorT2}\end{equation}
which can easily be checked to coincide with the general formula
(\ref{mirrormap}). The matrix (\ref{mirrorT2}) can likewise be written as in
(\ref{backgroundmatrix}) in terms of a mirror metric tensor and torsion form.

\newsubsection{Composite fermions and the Jain hierarchy}

Let us now turn to some basic ideas in the quantum Hall effect. Blok and Wen
\cite{blokwen} showed that the hierarchy scheme of states proposed by Jain
\cite{jain} for the fractional quantum Hall effect can be viewed in terms of an
effective theory of composite fermions. The basic idea of this contruction is
to consider a two-dimensional gas of electrons in an external magnetic field
$b$ at filling fraction
\begin{equation}
\nu_{\rm J}=\frac p{2mp+1}
\label{fillingfrac}\end{equation}
where $m$ and $p$ are integers, and attach $2m$ flux tubes to each electron.
The resulting gas of ``composite'' fermions (i.e. bound states of fermions and
flux tubes) carries an effective flux (and hence filling fraction)
$1/p=2m+1/\nu$. In this way, the $\nu=p/(2mp+1)$ fractional quantum Hall states
are related to the $\nu=p$ integer quantum Hall states of the composite object.

The effective action which describes the binding of $2m$ units of flux to each
electron is
\begin{equation}
S_{\rm{QH}}=\int_{\cal M}d^3x~\left(\psi^\dagger\,i\left(\partial_0-
iA_0-iea_0\right)\psi+\frac{1}{2m_e}\psi^\dagger\left(\nabla-iA-iea\right)^2
\psi\right)+\frac{1}{m}S_{\rm{CS}}^{[U(1)]}[A]
\end{equation}
where $\psi$ is the fermionic field for the composite object, $a$ is the
external electromagnetic vector potential and $A$ is a ``fictitious'' abelian
Chern-Simons gauge field. Here $e$ is the electron charge and $m_e$ the
electron mass. The equation of motion for $A_0$,
\begin{equation}
n_e\equiv
\left\langle\psi^\dagger\psi\right\rangle=-\frac{1}{2\pi}\frac{1}{2m}\langle
B\rangle
\end{equation}
implies that each electron carries $2m$ units of flux and so the fermions see
(on average) an effective magnetic field $B_{\rm{eff}}=b+\langle B\rangle$.
Since the filling fraction specifies the number of filled Landau levels,
$\nu\propto n_e/B$, this implies that the effective filling fraction of the
composite object moving in the effective magnetic field is
\begin{equation}
\frac{1}{\nu_{\rm{eff}}}=\frac{2mp+1}{p} - 2m=\frac{1}{p}
\end{equation}

The most general fractional quantum Hall states are described by the effective
theory
\begin{equation}
S[A,a]=2K\cdot S_{\rm CS}^{[U(1)^p]}[A]+\sum_I\int_{\cal M}\left(A^I\wedge\star
J_I+\frac{e{\cal Q}_I}{2\pi}\,a\wedge dA^I\right)
\label{qhe}\end{equation}
The state described by (\ref{qhe}) contains ${\rm rank}(K)$ quasi-particle
excitations which have current densities $J_I$. The index $I$ labels different
levels of the quasi-particles. The electromagnetic field $a$ minimally couples
to each of the topological currents $\star dA^I$ with charge strengths $e{\cal
Q}_I$. By integrating over $A^I$ one can prove that the filling fraction for
this state is given by
\begin{equation}
\nu=\mbox{$\sum_{I,J}$}\,{\cal Q}_I(K^{-1})^{IJ}{\cal Q}_J
\label{nuK}\end{equation}
Different hierarchy schemes of quantum Hall states correspond to different
forms of the Chern-Simons coefficient matrix $K$ and the charges ${\cal Q}_I$.
For the case of the Jain hierarchy, the filling fractions (\ref{fillingfrac})
are obtained by taking ${\cal Q}_I=1~~\forall I=1,\dots,p$ and the $p\times p$
matrix
\begin{equation}
K^{\rm (J)}=\pmatrix{1&0&\dots&0\cr0&1&\dots&0\cr &\dots&\dots&
\cr0&0&\dots&1\cr}+2m\pmatrix{1&1&\dots&1\cr1&1&\dots&1\cr &\dots&\dots&
\cr1&1&\dots&1\cr}
\label{jaink}\end{equation}
It is important to realize though that there can be many different fractional
quantum Hall states corresponding to the same filling fraction $\nu$. Such
states do, however, exhibit some degree of topological order which is
intimately related to the properties of the edge theory \cite{wenET}.

\newsubsection{The mirror Jain hierarchy}

We now consider the effects of mirror symmetry of the edge theory as a map on
quantum Hall systems in the bulk. In terms of the $XY$ model the mirror map
exchanges spin wave and magnetic vortex degrees of freedom along the $p$-th
direction of $\Gamma$ \cite{sath}. A similar monopole-instanton induced process
for multi-layer quantum Hall systems has been described in \cite{wenzee}, where
it was shown that when an electron hops from one layer to another the
corresponding currents are no longer separately conserved but change in a
manner similar to the mirror map described in this paper (indicating the
presence of a monopole-instanton). The interlayer electron hopping corresponds
to the instanton described by a monopole. The monopoles exchange the fluxes
associated with each layer, and they have also been shown to convert an anyon
superfluid into a normal fluid \cite{wenzee}. The role of duality in the
quantum Hall effect has also been discussed in \cite{shapwil} where it was
shown to be the transformation $\nu\to1-\nu$ that corresponds to particle-hole
conjugation with respect to the completely filled state. In \cite{bal} it was
shown how to use elements of the duality group of toroidally compactified
string theory to map some of the well-known hierarchies of quantum Hall states
into one another.

The effective edge theory for the quantum Hall system described by the bulk
action (\ref{qhe}) is identical to the action (\ref{XYaction}) for bosonic
string theory compactified on a $p$-dimensional torus. The mirror edge theory
itself can then be viewed as originating from some mirror three-dimensional
Chern-Simons gauge theory with coefficient matrix $\widetilde{K}$. The mirror
filling fraction for this new quantum Hall system is given by
\begin{equation}
\widetilde{\nu}=\mbox{$\sum_{I,J}$}\,{\cal
Q}_I(\widetilde{K}_{\rm{sym}}^{-1})^{IJ}{\cal Q}_J
\label{mfill}\end{equation}
Note that the mirror matrix $\widetilde{K}$ in (\ref{mirrormap}) will in
general contain an anti-symmetric part. This piece corresponds to a torsion
form on the boundary, and acts as a topological instanton term affecting only
the global, topological properties of the edge theory. Upon integrating
(\ref{csactionp}) by parts, we see that only the symmetric part of
$\widetilde{K}$ contributes to the bulk dynamics and hence to local observables
such as the Hall conductivity $\sigma_H$. Therefore, before inversion, we must
symmetrize $\widetilde{K}$ in order to calculate the mirror filling fraction
using (\ref{mfill}).

To see the effects of mirror symmetry on states of the Jain hierarchy, we use
the same charges ${\cal Q}_I=1$, i.e. all the Chern-Simons fields carry the
charge of an electron. Using (\ref{mirrormap}) and (\ref{jaink}) we find after
some algebra that
\begin{eqnarray}
\widetilde{K}^{\rm (J)}_{\rm
sym}&=&\frac1{2m+1}\pmatrix{4m+1&2m&\dots&2m&0\cr2m&4m+1&\dots&2m&0\cr
&\dots&\dots&\dots& \cr2m&2m&\dots&4m+1&0\cr0&0&\dots&0&1\cr}\nonumber
\\&=&\pmatrix{1&0&\dots&0&0\cr0&1&\dots&0&0\cr &\dots&\dots&\dots&
\cr0&0&\dots&1&0\cr0&0&\dots&0&\mbox{$\frac1{2m+1}$}\cr}+\frac{2m}{2m+1}
\pmatrix{1&1&\dots&1&0\cr1&1&\dots&1&0\cr &\dots&\dots&\dots&
\cr1&1&\dots&1&0\cr0&0&\dots&0&0\cr}
\label{mirrorjaink}\end{eqnarray}
Inverting (\ref{mirrorjaink}) we find that the mirror transformation maps the
filling fractions as
\begin{equation}
\nu_{\rm J}=\frac{p}{2mp+1}~\longrightarrow~\widetilde{\nu}_{\rm
J}=2(m+1)+\frac{p-2(m+1)}{2mp+1}=(2m+1)^2\nu_{\rm J}
\label{fillingtransf}\end{equation}
Hence the mirror quantum Hall system can be interpreted as having $2(m+1)$
completely filled Landau levels plus $p-2(m+1)$ partially filled levels. Since
the number of independent degrees of freedom in the Jain picture can be
calculated as the sum of the number of filled Landau levels and the number of
partially filled levels, we see that both the original and the mirror quantum
Hall systems have $p={\rm{rank}}(K)$ degrees of freedom.

The mirror system defined by the coefficient matrix (\ref{mirrorjaink})
represents a new hierarchy of states of the quantum Hall effect. Note that for
$m=-1$, the mirror map leaves the filling fraction (\ref{fillingtransf})
invariant. Thus the lowest lying composites of the Jain hierarchy are mapped
into the {\it same} (class of) quantum Hall systems under mirror symmetry, with
a new hierarchical description determined by (\ref{mirrorjaink}). For generic
$m$, it is intriguing to interpret the mirror filling fraction using a
composite fermion picture by writing
\begin{equation}
\frac{1}{\widetilde{\nu}_{\rm J}}=\frac{1}{p(2m+1)^2}+\frac{2m}{(2m+1)^2}
\end{equation}
This shows that in the mirror theory there are $2m/(2m+1)^2$ fractional units
of flux attached to each electron and so the quasi-particles become anyons
which see an effective magnetic field $\widetilde{B}_{\rm eff}=B_{\rm
eff}/(2m+1)^2$. Thus we have an explicit physical realization of the anyonic
magnetic flux symmetry responsible for the mirror map. Since the number of
filled Landau levels is inversely proportional to the effective magnetic field
seen by the electrons, this yields a physical reason as to why mirror symmetry
maps a quantum Hall system with a partially-filled Landau level to another
quantum Hall system with more than one filled Landau level.

The mirror map is not a symmetry of the edge theory induced by the bulk action
(\ref{qhe}), because of the gauging of the matter-coupled Chern-Simons terms
with respect to the electromagnetic field. However, the spectra of the edge
excitations are identical for the Hall conductivity $\sigma_H\propto\nu$ and
its mirror. For the Jain hierarchy, the only true symmetry appears at the point
$m=-1$ in the moduli space. What we have here is a non-perturbative, topology
changing mechanism that maps a given quantum Hall system onto another, thus
determining new hierarchies of states. Combinations of these mappings could
eventually leave the filling fractions determined as in (\ref{fillingtransf})
invariant, leading to new sorts of topological symmetries of the quantum Hall
effect. In this way the role of the monopole-instantons seems to affect the
quasi-particle excitations in some non-trivial way, for example turning
composite fermions into anyons, and interchanging winding modes of the
quasi-particles with magnetic monopole configurations of the Chern-Simons gauge
fields both of which interact with the electromagnetic field. It remains to
determine precisely how all of these topological effects affect the global and
local properties of quantum Hall systems.

\newsection{Conclusions}

In this paper we have described mirror transformations in three theories.
Beginning with an algebraic isomorphism of $N=2$ superconformal field theories,
we described how two topologically distinct Calabi-Yau manifolds were identical
from the point of view of quantum string theory. We then showed how two
topologically inequivalent topological membranes induce the same string theory,
and how all of the basic topological symmetries of the target space are
realized as discrete geometry altering transformations of a 3-manifold $\cal
M$. The mirror map in this case is non-perturbative in character and it
interchanges particle winding numbers with monopole numbers. This suggests that
mirror symmetry in this description is a sort of $S$-duality. We also showed
that, in the three-dimensional picture, there is an intimate connection between
mirror symmetry and $T$-duality. This confirms other expectations about the
realization of the mirror transformation from a $T$-dual perspective, such as
the proposal of \cite{strom} that mirror symmetry can be thought of as
$T$-duality ($\star d\leftrightarrow d$) on toroidal fibers which are
sypersymmetric 3-cycles of certain Calabi-Yau spaces, or the more recent
algebraic realization of the action of mirror symmetry as a Poisson-Lie
$T$-duality transformation of the $N=2$ superconformal algebra \cite{park}. It
is intriguing to note that the major role played by monopole-instantons in the
three-dimensional picture is reminescent of the role of instantons (complex
curves) in Witten's linear sigma-model approach to quantum geometry
\cite{witteninst}. Moreover, the map between particle winding modes and
monopole-instantons could be a link between topological membrane theory and
11-dimensional M Theory, in light of the recent realization that the
interchange of momentum modes with instantons is a symmetry of Matrix Theory
(see \cite{matrix} and references therein).

One important aspect that we have not described in this paper is how to
explicitly obtain other conventional realizations of the $N=2$ superconformal
algebra, such as those provided by Landau-Ginsburg orbifold models. These
models are important for the description of flows on the moduli space of $N=2$
superconformal field theories, and hence for more complete descriptions of
mirror manifolds. These potentially complicated processes are presumably
described by charge deformations arising from the coupling of Chern-Simons
gauge theories to {\it dynamical} charged matter fields. The study of these
more involved dynamical models could relate the geometric operation of the
mirror map in terms of topology changing processes on principal fiber bundles
to those predicted from algebraic geometry, such as the `flop' transition in
which the area of a rational curve is shrunk down to zero size and then
expanded back to positive volume in a transverse direction
\cite{aspinwall,greene}, and also from the combinatorical ideas of toric
geometry \cite{greene}. This could also lead to three-dimensional realizations
of spacetime topology change from conifold singularities in which physically
smooth transitions between Calabi-Yau manifolds with {\it different} Hodge
numbers occur \cite{gms}. This would link the properties of superstrings, black
holes and $D$-branes to the topological membrane approach to string theory.

The third model in which we described mirror symmetry was the mean field theory
for the quantum Hall effect. We showed how it implies a non-trivial
hierarchical changing process in these systems, although the (physical)
interpretation of the mirror map in these cases is far less clear. It would be
interesting to analyse more carefully the physics associated with the mirror
Hall systems, such as the direct interpretation of the magnetic
monopole-instantons in the corresponding physical states, and to determine if
such systems are indeed experimentally observable. This would then imply the
existence of an experimental laboratory in which one could study the physics of
mirror symmetry.

\vspace{1cm}

\noindent
{\bf Acknowledgements:} We thank A. Lopez for many helpful discussions
concerning the quantum Hall effect. {\sc l.c.} gratefully acknowledges
financial support from the University of Canterbury, New Zealand. The work of
{\sc i.i.k.} and {\sc r.j.s.} was supported in part by the Particle Physics and
Astronomy Research Council (U.K.).


\newpage

\begin{thebibliography}{99}

\baselineskip=12pt

\bibitem{mirror} L. Dixon, in: {\em Superstrings, Unified Theories and
Cosmology 1987}, eds. G. Furlan et al. (World Scientific, 1988);\\ W. Lerche,
C. Vafa and N. Warner, Nucl. Phys. B 324 (1989) 427.

\bibitem{mirrors} B.R. Greene and M.R. Plesser, Nucl. Phys. B 338 (1990) 15;\\
P. Candelas, P.S. Green and T. H\"ubsch, Nucl. Phys. B 341 (1990) 383.

\bibitem{aspinwall} P.S. Aspinwall, in: Proc. Trieste Summer School in High
Energy Physics, eds. E. Gava, A. Masiero, K.S. Narain, S. Randjbar-Daemi and Q.
Shafi (World Scientific, 1995).

\bibitem{mirror1} S.-T. Yau, ed., {\it Essays on Mirror Manifolds}
(International Press, 1992).

\bibitem{mirror2} B.R. Greene and S.-T. Yau, eds., {\it Essays in Mirror
Manifolds II} (International Press, 1995).

\bibitem{greene} B.R. Greene, {\em String Theory on Calabi-Yau
Manifolds}, hep-th/9702155.

\bibitem{witten} E. Witten, Commun. Math. Phys. 121 (1989) 351.

\bibitem{mooreET} G. Moore and N. Seiberg, Phys. Lett. B 220 (1989) 422.

\bibitem{tm} I.I. Kogan, Phys. Lett. B 231 (1989) 377;\\ S. Carlip and I.I.
Kogan, Phys. Rev. Lett. 64 (1990) 148; Mod. Phys. Lett. A 6 (1991) 171.

\bibitem{szaboET} G. Amelino-Camelia, I.I. Kogan and R.J. Szabo, Nucl. Phys. B
480 (1996) 413.

\bibitem{KZ} V.G. Knizhnik and A.B. Zamolodchikov, Nucl. Phys. B 247 (1984) 83.

\bibitem{anyon} F. Wilczek, Phys. Rev. Lett. 48 (1982) 1144, 1146; 49 (1982)
957;\\ F. Wilczek and A. Zee, Phys. Rev. Lett. 51 (1983) 2250;\\ D. Arovas, R.
Schrieffer, F. Wilczek and A. Zee, Nucl. Phys. B 251 [FS13] (1985) 117;\\ C.-H.
Tze and S. Nam, Ann. Phys. 193 (1989) 419.

\bibitem{us} L. Cooper, I.I. Kogan and R.J. Szabo, Nucl. Phys. B 498 (1997)
492.

\bibitem{qhecs} G.W. Semenoff and P. Sodano, Phys. Rev. Lett. 57 (1986) 1195;\\
R. Prange and S. Girvin, eds., {\it The Quantum Hall Effect} (Springer, Berlin,
1987);\\ T.H. Haansson, S. Kivelson and T.H. Zhang, Phys. Rev. Lett. 62 (1989)
82.

\bibitem{cardy} A.B. Zamolodchikov, JETP Lett. 43 (1986) 565; Sov. J. Nucl.
Phys. 46 (1987) 1090;\\ A.A.W. Ludwig and J.L. Cardy, Nucl. Phys. B 285 (1987)
687.

\bibitem{witteninst} E. Witten, Nucl. Phys. B 403 (1993) 159; in
\cite{mirror2};\\ P.S. Aspinwall, B.R. Greene and D.R. Morrison, Nucl. Phys. B
416 (1994) 414.

\bibitem{minmirror} B.R. Greene, C. Vafa and N. Warner, Nucl. Phys. B 324
(1989) 371;\\ E. Martinec, Phys. Lett. B 217 (1989) 431.

\bibitem{zamfat} A.B. Zamolodchikov and V.A. Fateev, Sov. Phys. JETP 63 (1986)
91.

\bibitem{gepnerET1987a} D. Gepner and Z. Qiu, Nucl. Phys. B 285 [FS19] (1987)
423.

\bibitem{kogan2} I.I. Kogan, Mod. Phys. Lett. A 6 (1991) 501.

\bibitem{kogan3} I.I. Kogan, Phys. Lett. B 255 (1991) 31.

\bibitem{kogan1} I.I. Kogan, Phys. Lett. B 390 (1997) 189.

\bibitem{greene1996a} B.R. Greene, in \cite{mirror2}.

\bibitem{monoinst} M. L\"{u}scher, Nucl. Phys. B 326 (1989) 557;\\ K. Lee,
Nucl. Phys. B 373 (1992) 735;\\ I.I. Kogan and A. Kovner, Phys. Rev. D 53
(1996) 4510.

\bibitem{kaiming} L. Cooper, I.I. Kogan and K.-M. Lee, Phys. Lett. B 394 (1997)
67.

\bibitem{bergeron} M. Bos and V.P. Nair, Phys. Lett. B 223 (1989) 61;\\ M.
Bergeron, D. Eliezer and G.W. Semenoff, Phys. Lett. B 311 (1993) 137;\\ M.
Bergeron and G.W. Semenoff, Ann. Phys. 245 (1996) 1.

\bibitem{tmgt} J.F. Schonfeld, Nucl. Phys. B 185 (1981) 157;\\ S. Deser, R.
Jackiw and S. Templeton, Ann. Phys. 140 (1982) 372.

\bibitem{heterotic} L. Cooper and I.I. Kogan, Phys. Lett. B 383 (1996) 271.

\bibitem{gepner} D. Gepner, Phys. Lett. B 199 (1987) 380; Nucl. Phys. B 296
(1988) 757.

\bibitem{emss} S. Elitzur, G. Moore, A. Schwimmer and N. Seiberg, Nucl. Phys. B
326 (1989) 108.

\bibitem{procatmgt} S. Deser and R. Jackiw, Phys. Lett. B 139 (1984) 371;\\ A.
Karlhede, U. Lindstr\"om, M. Rocek and P. van Nieuwenhuizen, Phys. Lett. B 186
(1987) 96.

\bibitem{polyakov1988} A.M. Polyakov, Mod. Phys. Lett. A 3 (1988) 325.

\bibitem{sath} B. Sathiapalan, Phys. Rev. D 35 (1987) 3277;\\ I.I. Kogan, JETP
Lett. 45 (1987) 709;\\ A.A. Abrikosov, Jr. and I.I. Kogan, Sov. Phys. JETP 96
(1989) 418.

\bibitem{giveon} A. Giveon, M. Porrati and E. Rabinovici, Phys. Rep. 244 (1994)
77.

\bibitem{blokwen} B. Blok and X.G. Wen, Phys. Rev. B 42 (1990) 8133.

\bibitem{jain} J.K. Jain, Phys. Rev. Lett. 63 (1989) 199; Phys. Rev. B 40
(1989) 8079.

\bibitem{wenET} X.G. Wen, Phys. Rev. B 40 (1989) 7378; Intern. J. Mod. Phys. B
4 (1990) 239;\\ X.G. Wen and Q. Niu, Phys. Rev. B 41 (1990) 9377.

\bibitem{wenzee} X.G. Wen and A. Zee, Phys. Rev. B 41 (1990) 240; Phys. Rev.
Lett. 69 (1992) 1811.

\bibitem{shapwil} A. Shapere and F. Wilczek, Nucl. Phys. B 320 (1989) 669.

\bibitem{bal} A.P. Balachandran, L. Chandar and B. Sathiapalan, Nucl. Phys. B
443 (1995) 465; Intern. J. Mod. Phys. A 11 (1996) 3587.

\bibitem{strom} A. Strominger, S.-T. Yau and E. Zaslow, Nucl. Phys. B 479
(1996) 243.

\bibitem{park} S.E. Parkhomenko, {\it Mirror Symmetry as a Poisson-Lie
$T$-duality}, hep-th/9710037.

\bibitem{matrix} R. Gopakumar, Nucl. Phys. B 507 (1997) 609;\\ J. Polchinski
and P. Pouliot, Phys. Rev. D 56 (1997) 6601.

\bibitem{gms} B.R. Greene, D.R. Morrison and A. Strominger, Nucl. Phys. B 451
(1995) 109.

\end{thebibliography}
\end{document}